\documentclass[a4paper,11pt]{article}
\pdfoutput=1 %

\usepackage{jheppub}
\usepackage[T1]{fontenc} 
\usepackage[normalem]{ulem}
\usepackage{slashed}
\usepackage{placeins} 
\usepackage{comment}
\usepackage{amsmath,amssymb,graphicx,multirow,xspace,slashed,subfigure}

\usepackage{appendix}

\usepackage{xcolor}

\definecolor{darkgreen}{rgb}{0,0.5,0}
\definecolor{goodyellow}{rgb}{0.9,0.7,0}

\newcommand\Eq[1]{Eq.~\ref{#1}}
\newcommand\Fig[1]{Fig.~\ref{#1}}
\newcommand\Sec[1]{Sec.~\ref{#1}}

\newcommand{\pt}{\ensuremath{p_T}}

\newcommand{\pbb}{$\mathcal{M}_2$}
\newcommand{\un}{$\mathcal{U}_{N}$}
\newcommand{\evCol}{$\mathcal{D}_{\mathcal{M}_2}$}
\newcommand{\evIso}{$\bar{\mathcal{I}}$}

\preprint{MIT-CTP/5755}

\title{A Field Guide to Event-Shape Observables Using Optimal Transport}

\author{Cari Cesarotti$^{1,2}$,}
\author{Matt LeBlanc$^{3,2}$}

\affiliation{$^1$Center for Theoretical Physics, Massachusetts Institute of Technology, Cambridge, MA, USA}
\affiliation{$^2$The NSF AI Institute for Artificial Intelligence \& Fundamental Interactions, Boston, MA, USA}
\affiliation{$^3$Department of Physics, Brown University, Providence, RI, USA}

\emailAdd{ccesar@mit.edu}
\emailAdd{matt.leblanc@cern.ch}

\abstract{%
We lay out the phenomenological behavior of event-shape observables evaluated by solving optimal transport problems between collider events and reference geometries---which we name \textit{manifold distances}---to provide guidance regarding their use in future studies.
This discussion considers several choices related to the metric used to quantify  these distances.
We explore the differences between the various options, for the first time using a combination of analytical studies and simulated minimum-bias and multi-jet events.
Making judicious choices when defining the  metric and reference geometry can improve sensitivity to interesting signal features and reduce sensitivity to non-perturbative effects in QCD.
The goal of this article is to provide a `field guide' that can inform how choices made when defining a manifold distance can be tailored for the analysis at-hand.
}

\begin{document}
\maketitle
\flushbottom

\section{Introduction}
Event-shape observables~\cite{Banfi:2010xy} quantify the flow of hadronic energy in high-energy collider events, like those produced at the Large Hadron Collider (LHC) at CERN.
They have repeatedly proven to be successful tools for understanding quantum chromodynamics (QCD); for example, event shapes were utilised in the experimental observation of the gluon~\cite{TASSO:1979zyf, BARTEL1980142, TASSO:1983cre}, have been applied in determinations of the strong coupling $\alpha_s$~\cite{BARBER1979139, L3:1992btq, SLD:1994idb, ALEPH:1995njx, 
OPAL:2001klt, ALEPH:2003obs, L3:2004cdh, OPAL:2004wof, H1:2005zsk, ZEUS:2006vwm, Kluth:2009nx, OPAL:2011aa, Abbiendi:2008zz, Nason:2023asn} and other properties of QCD from experimental data~\cite{Kluth:2000km,OPAL:2001klt,DELPHI:2003yqh,Dasgupta:2003iq}, and are invaluable tools for Parton Shower Monte Carlo (MC) development and tuning~\cite{MARK-II:1989aml,L3:1990jlf,TASSO:1990cdg,Skands:2014pea,ATL-PHYS-PUB-2014-021,CMS-GEN-17-001}.
The utility of event shapes also extends to searches for new physics, as these observables have also been shown to be effective at isolating signal-enhanced event geometries in a wide variety of beyond-the-Standard-Model (BSM) physics scenarios~\cite{Krauss:2016ely,Goldouzian:2020wdq,Buckley:2020wzk,Buckley:2021neu,Knapen:2016hky,Cesarotti:2020uod,Khoze:2019jta,Amoroso:2020zrz,Arkani-Hamed:1998jmv,Antoniadis:1998ig,Arkani-Hamed:1998sfv,Randall:1999ee,Randall:1999vf,Giddings:2001bu,Anchordoqui:2002fc,Landsberg:2006mm,Meade:2007sz,Dimopoulos:2001hw,Hirschi:2018etq,Dixon:1993xd,CMS-EXO-15-007}.

Event-shape observables are not universally discriminating; the choice of most effective observable is determined by the physics of interest to the analysis at-hand. 
For example, \textit{thrust}~\cite{PhysRevLett.39.1587,BRANDT196457,DERUJULA1978387} is often the preferred observable for analyses of QCD at high momentum transfers.
Thrust is a motivated choice in this case, as it quantifies how similar an event is to a $2\rightarrow2$ point-like particle scattering geometry. 
However, for physics scenarios with non-jetty topologies (such as homogeneous/uniform radiation patterns, high jet multiplicities, etc., thrust can be less effective for selecting signal events than other options.
These kinds of radiation patterns are created by collider-produced black holes~\cite{Giddings:2001bu, Dimopoulos:2001hw, Meade:2007sz}, multi-jet final states in supersymmetric scenarios~\cite{Evans:2013jna,Barbier:2004ez}, high-multiplicity and strongly coupled new physics scenarios \cite{Strassler:2006im, Polchinski:2002jw, Hofman:2008ar, Strassler:2008fv, Strassler:2008bv, Hatta:2008tx}, and models that generically fall under the category of soft, unclustered energy patterns (SUEPs)~\cite{Knapen:2016hky}.

A broad library of event shapes can facilitate investigations into diverse event geometries, helping to ensure that no stone is left unturned in the search for BSM physics at the LHC and at future colliders~\cite{Narain:2022qud}.
In this work, we examine the phenomenology of event-shape observables that are defined using a recently-proposed methodology of quantifying the distance of collider events to symmetric reference geometries in terms of a formal metric.
These observables are computed using the particle physics application of the Earth Mover's Distance~\cite{10.1007/978-3-642-40020-9_43,192468,Rubner:1998:MDA:938978.939133,Rubner2000,10.1007/978-3-540-88690-7_37} -- or Wasserstein distance~\cite{wasserstein1969markov, dobrushin1970prescribing} -- called the \textit{Energy Mover's Distance} (EMD) \cite{Komiske:2019fks,Komiske:2020qhg}.
We will generically refer to these observables as \textit{manifold distances}.
The first application of the EMD in the context of designing new event shapes was the definition of \textit{event isotropy} \cite{Cesarotti:2020hwb}, where the reference geometry is a (quasi-) isotropic radiation pattern.
Canonical event shapes like thrust, sphericity~\cite{PhysRevD.1.1416,Ellis:1976uc} and spherocity~\cite{PhysRevLett.39.1237} were shown to be less efficient discriminants for signals with quasi-isotropic signatures~\cite{Cesarotti:2020ngq}.
Event isotropy was recently measured in proton-proton collisions at the Large Hadron Collider in high-\pt{} multijet events by the ATLAS Collaboration~\cite{STDM-2020-20}, and in minimum-bias events by the CMS Collaboration~\cite{CMS:2024bwe}.
Additionally the EMD has been used to study jet substructure by proposing novel observables~\cite{Ba:2023hix, Kitouni:2022qyr}, and optimal transport more generally has been proposed to further study collider events and identify anomalies in the context of unsupervised learning studies~\cite{Cai:2020vzx, Cai:2021hnn} and novel triggering strategies~\cite{Craig:2024rlv}\footnote{Reviews of jet substructure and anomaly detection studies at the LHC may be found respectively in Refs.~\cite{Kogler:2018hem} and~\cite{Kasieczka:2021xcg} for readers unfamiliar with these subfields.}.

Here, we build on the foundation laid by event isotropy and address questions relevant for more generalized event shape definitions using EMDs. 
We do so by varying the two fundamental components that specify the observable: the reference geometry and the distance metric.\footnote{One could also change the energy dependence, which is related to changing the power in computing in the sphericity tensor \cite{PhysRevD.1.1416, Ellis:1976uc} or the $p_T^D$ observable \cite{CMS-HIG-11-027}. As this change would result in the loss of IRC safety, we have not explored this topic in this work.}
We consider two complementary geometries in the transverse plane: an $N=2$ back-to-back particle event $\mathcal{E}_2$ from the manifold of such events \pbb, and a discrete, uniform `ring-like' radiation pattern \un.
The back-to-back configuration was first mentioned and computed by ATLAS in Ref.~\cite{STDM-2020-20}, but never in earlier phenomenological studies. 
Additionally, we consider several different distance metrics to illustrate how the choice of metric can affect the sensitivity of observables that we construct to distinct QCD physics processes.

The outline of the paper is as follows: in Section~\ref{sec:emd}, a review of the EMD and its use in defining event-shape observables is provided along with discussion related to relevant analytics for observable calculation.
The main results and discussion is found in Section~\ref{sec:results}, where we show various 1- and 2-dimensional distributions to discuss how the observables behave, when they are correlated, and how the choices made when defining these observables yields sensitivity to different features.
Finally, some concluding remarks and prescriptions for future analyses are offered in Section~\ref{sec:conclusions}.

\section{Definition of event observables with EMD}
\label{sec:emd}
To begin, we review how observables are constructed and computed with the EMD, which provides a well-defined distance metric between two distributions of final-state radiation patterns\footnote{In this work, we take an event to be synonymous with its radiation pattern, although the objects in the event (\emph{e.g.} stable hadrons, partons, jets, \emph{etc.}) will vary.}.
More concretely, it is the minimum work needed to rearrange one set of (particle) four-momenta into a second set. 
While the EMD was first introduced to measure the distance between two events, subsequent work has shown that the EMD can also be used quantify the radiation pattern of individual events to better understand the underlying physics \cite{Cesarotti:2020hwb, Cesarotti:2020ngq}. 
Rather than comparing an event with other events, the event is instead compared to a template radiation pattern that is called the \textit{reference geometry}.
The details of how these geometries are defined are given in Ref.~\cite{Cesarotti:2020hwb, Komiske:2020qhg, Ba:2023hix}; we summarize the central ideas here for completeness.

To construct a manifold distance, we compute the EMD in the simplified scenario of two events $\mathcal{E}$, $\mathcal{G}$ of equal energy with $i$ and $j$ massless particles $p_i$ and $p_j$, respectively.
$\mathcal{E}$ represents the collider event, and $\mathcal{G}$ the reference geometry. 
The EMD between $\mathcal{E}$ and $\mathcal{G}$ is defined as 
\begin{equation}
    \text{EMD}(\mathcal{E}, \mathcal{G}) \equiv \underset{\{f_{ij}\}}{\min}\sum_{ij} f_{ij} d_{ij} 
\label{eq:EMDdef}
\end{equation}
where the transport map $f_{ij}$ specifies how much energy is moved from $p_i \in \mathcal{E}$ to $p_j \in \mathcal{G}$, where 
\begin{equation}
\begin{split}
   & f_{ij} \geq 0 \\
   & \sum_i f_{ij} = E_j, \quad  \sum_j f_{ij} = E_i, \quad \sum_{ij} f_{ij} = E_\text{tot}
\end{split}
\label{eq:fijreq}
\end{equation}
for $E_i$ the energy weight of particle $p_i$, and $d_{ij}$ is the ground-space distance between particles. 
Both the energy weight and ground distance measure are user-defined inputs of the calculation and should be chosen to reflect the experimental set-up.
For consistent results, the inter-particle distance must be monotonic.

\begin{figure}
\begin{center}
\includegraphics[width=0.5\textwidth]{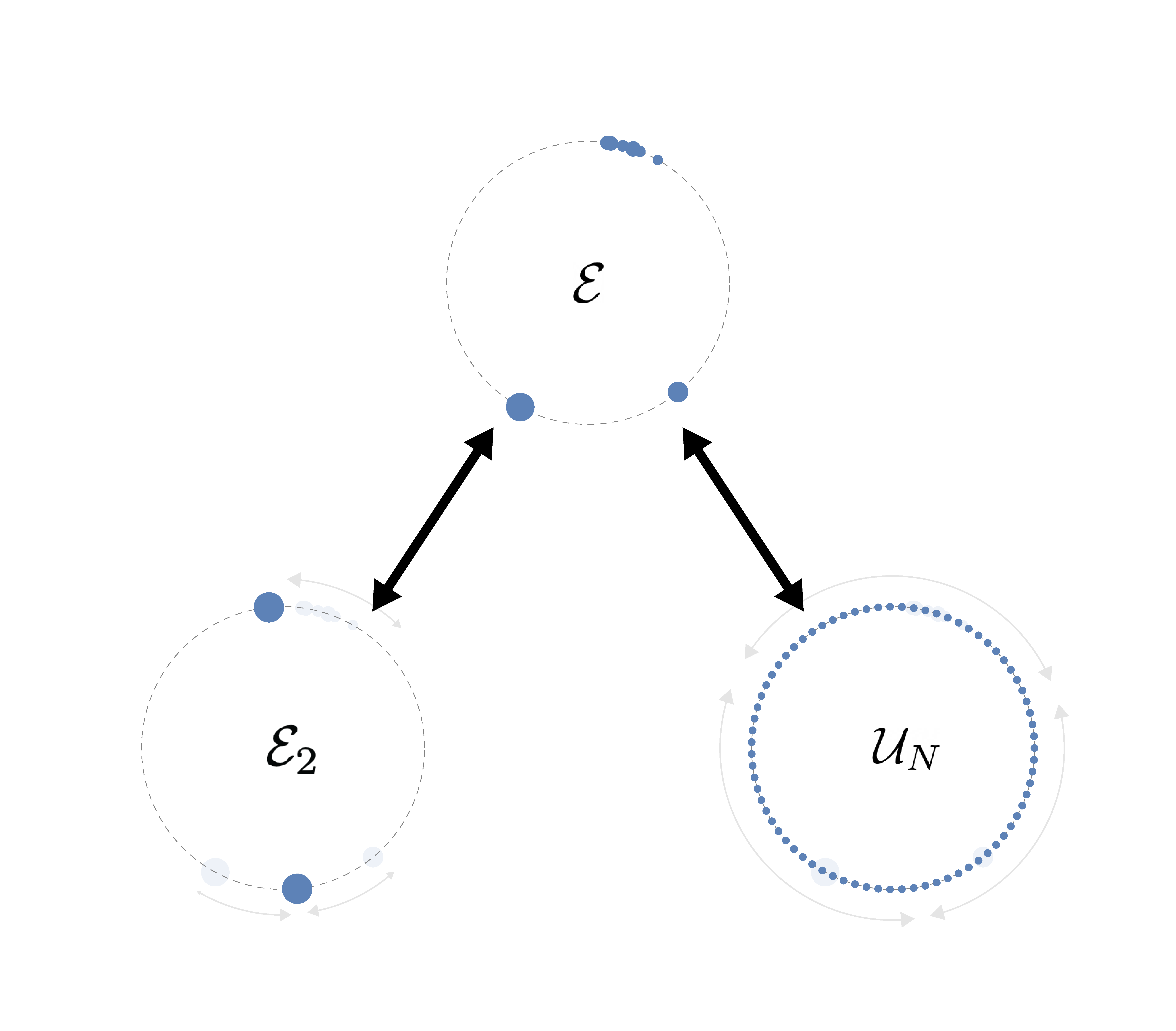}
\caption{A cartoon of the EMD as an event-shape observable.
The transverse projection of a collider event $\mathcal{E}$ is drawn such that the weight of the marker corresponds to the particle $p_T$. 
Some examples of reference geometries to compare against are two particle back-to-back configurations (also called \textit{dipole}) $\mathcal{E}_2$ from the manifold \pbb~and the quasi-uniform $\mathcal{U}_N$. 
One can imagine the EMD as the minimal work done to spread event $\mathcal{E}$ into either $\mathcal{E}_2$ or $\mathcal{U}_N$.}
\label{fig:emdIll}
\end{center}
\end{figure}

This construction can be generalized to probe other global event geometries or specific structures by comparing the collider event to reference geometries.
While there are many interesting candidates, we focus on two extremal examples: isotropic and dipole-like (back-to-back), as illustrated in Fig.~\ref{fig:emdIll}. 
In the original work, the reference geometry was chosen to be a uniform (or, quasi-uniform, for computational tractability) distribution \cite{Cesarotti:2020hwb}, but we will also consider the dipole geometry that was first shown in Ref.~\cite{STDM-2020-20} to be an observable comparable to thrust.
Additionally, we consider several distance measures to demonstrate how the sensitivity of these observables can be tuned to identify specific features of a geometry, \emph{e.g.} by only considering transverse information in hadron collisions due to the unknown boost along the beamline.\footnote{For the interested reader: Ref.~\cite{STDM-2020-20} conducts some studies with longitudinal information included.}

\subsection{Reference Geometries}
\label{sec:RefGeo}
In this section we define the two primary reference geometries designed to probe complementary radiation patterns: isotropic and collimated. 
We characterize the observable by the reference geometry.
For any reference geometry, the notation we adapt for the manifold distance of an event $\mathcal{E}$ to an event $\mathcal{E}'$ in in the manifold $\mathcal{M}$ such that the EMD is minimized is 
\begin{equation}
    \mathcal{D}_{\mathcal{M}}(\mathcal{E}) \equiv C \times \min_{\mathcal{E}'}\text{EMD}(\mathcal{E}, \mathcal{E}') \qquad \mathcal{E}' \in \mathcal{M}
\end{equation}
where $C$ is some constant for normalization.
For the isotropic sample, we keep the previously introduced name of \textit{event isotropy} $\mathcal{I}$.
This is a unique example of the manifold distance because the reference geometry $\mathcal{U}$ is perfectly uniform, and therefore its orientation does not need to be optimized\footnote{In other words: the manifold of perfectly-uniform radiation patterns in the transverse plane consists of only $\mathcal{U}$.}.
Alternatively, we consider what we call the \textit {dipole manifold distance} (\textit{dipole distance} for brevity), computed with a specific back-to-back two particle event $\mathcal{E}_2$ within the manifold of \pbb~that minimizes the EMD.
%

\subsubsection*{Isotropic reference}
The first geometry of interest is a uniform distribution of particles in a ring, introduced in Ref.~\cite{Cesarotti:2020hwb}. 
This observable is particularly motivated for use in searches for new physics~\cite{Giddings:2001bu, Dimopoulos:2001hw, Meade:2007sz,Evans:2013jna,Barbier:2004ez,Strassler:2006im, Polchinski:2002jw, Hofman:2008ar, Strassler:2008fv, Strassler:2008bv, Hatta:2008tx,Knapen:2016hky}, as isotropic radiation patterns are inherently suppressed in background QCD processes.
Thus, we expect an observable that quantifies the isotropy of an event would be a good discriminator between signal and background of standard QCD dijet events.

\begin{figure}
\begin{center}
\includegraphics[width=0.25\textwidth]{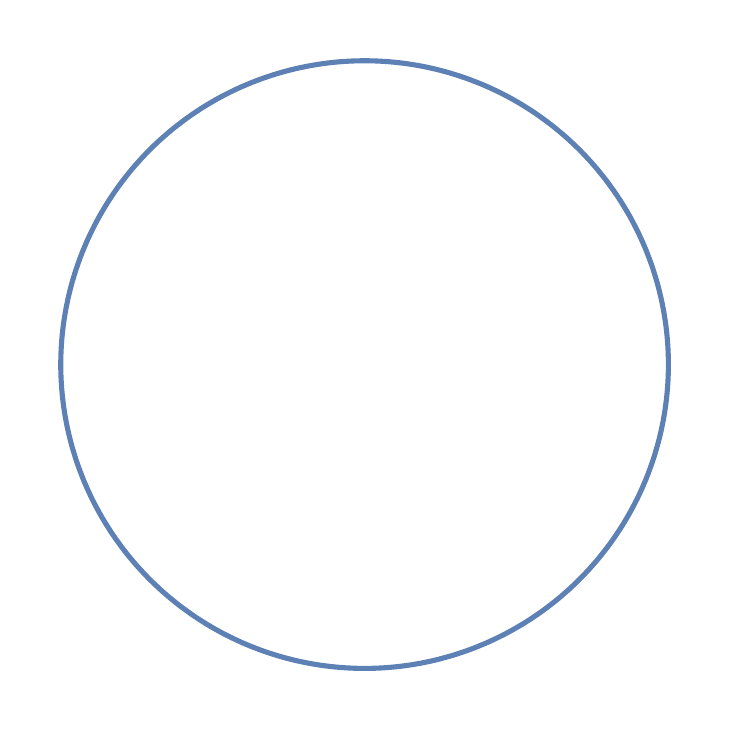} $\mathcal{U}$
\qquad \qquad 
\includegraphics[width=0.25\textwidth]{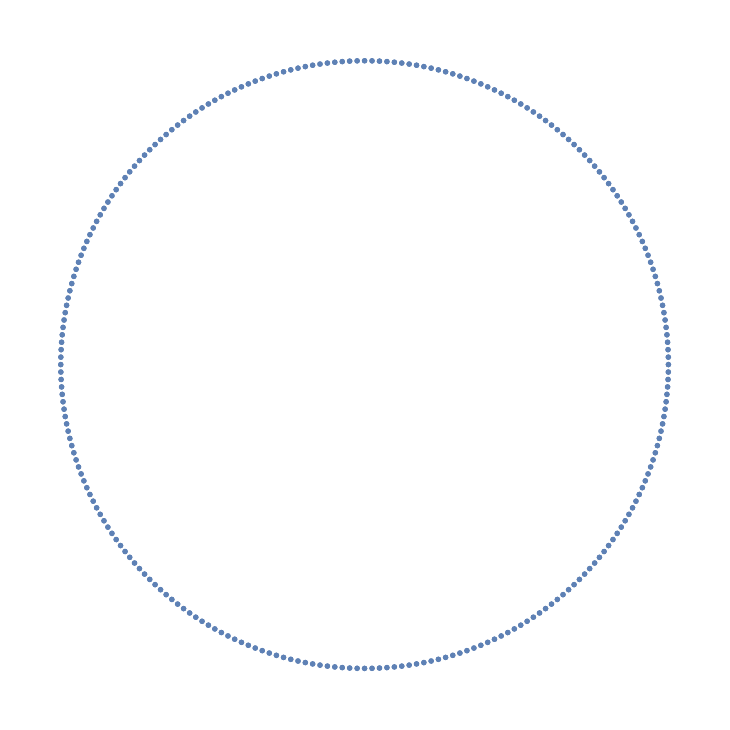}
$\mathcal{U}_{N=256}$
\end{center}
\caption{A perfectly uniform event (left) compared to the discretization into 256 particles used for computation (right).}
\label{fig:uniform}
\end{figure}

We define an isotropic event $\mathcal{U}$ as a radiation pattern with momentum uniformly distributed in the transverse plane,
\begin{equation}
\mathcal{U} = \left\{ E_\text{tot} \frac{d\phi}{2 \pi} (\cos\phi, \sin\phi) \ \big | \ \phi \in [0, 2\pi] \right\},
\label{eq:unifP}
\end{equation}
which we interpret as an infinite set of particles with infinitesimal momentum $d p_T = d\phi/2\pi$. 
While this is an interesting theoretical limit for analytical estimates, it is intractable for numeric calculations. 
To compute the EMD we use a quasi-isotropic radiation pattern with $N$ particles evenly distributed along the ring,
\begin{equation}
\mathcal{U}_N = \left\{ \frac{ E_\text{tot}}{N}\left(\cos\frac{2\pi n}{N}, \sin\frac{2\pi n}{N}\right) \ \big | \ n = 1, 2, ..., N \right\},
\label{eq:uniFinite}
\end{equation}
where $N$ is sufficiently large to approximate the infinite particle distribution, such that random rotations on the order of $\Delta \phi \sim 2 \pi / N$ do not effect the distribution of the observable.
For this work, we use $N=256$, such that $\mathcal{U} \approx \mathcal{U}_{256}$, shown in Fig.~\ref{fig:uniform}. 
We suppress these subscripts for notational simplicity, and will refer to the quasi-isotropic event as $\mathcal{U}$. 
%
\subsubsection*{Back-to-back reference}
The other geometry of interest is in the opposite regime: the equal energy, two-particle back-to-back configuration.
An observable utilizing this reference geometry was first introduced in Ref.~\cite{STDM-2020-20}.
We call this new observable \emph{dipole manifold distance}, and will abbreviate it as dipole distance or just \evCol~in plots.
As discussed in Ref.~\cite{Komiske:2020qhg} it is similar, but not identical, to thrust.
For an in-depth comparison of thrust to dipole distance, see Appendix \ref{app:thrust} of this document. 
\begin{figure}[t!]
\begin{center}
\includegraphics[width=0.25\textwidth]{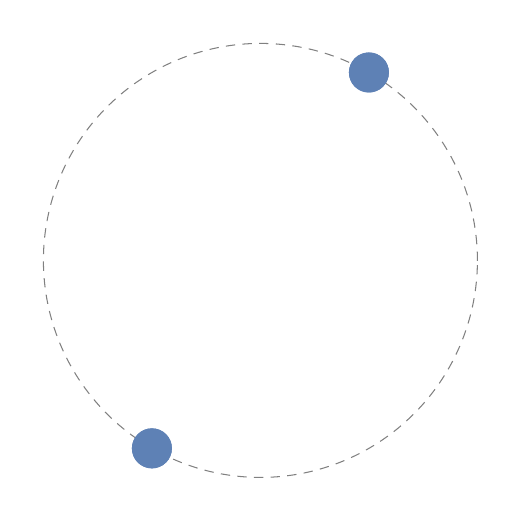} 
\qquad
\includegraphics[width=0.25\textwidth]{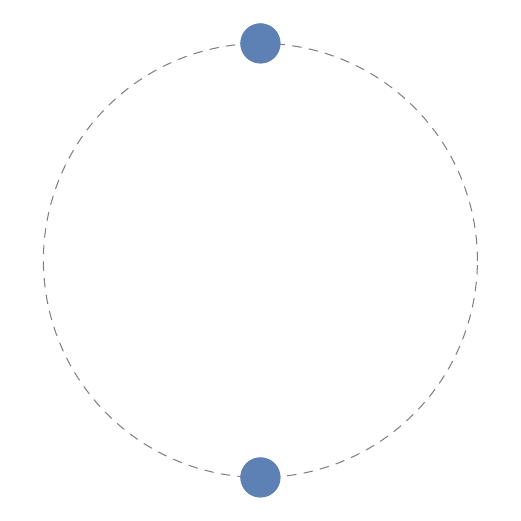}
\qquad
\includegraphics[width=0.25\textwidth]{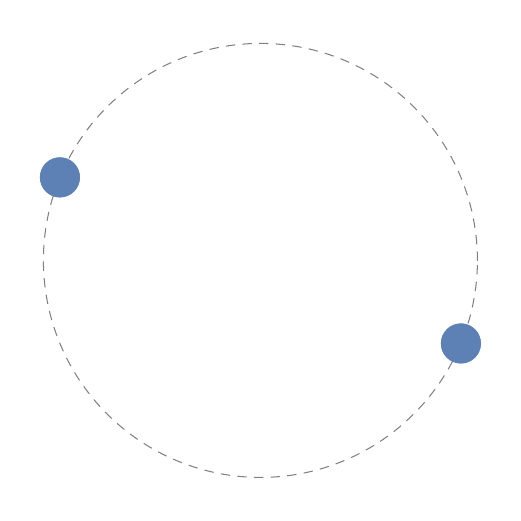}
\end{center}
\caption{An example of several possible dipole events in \pbb. The angle of orientation is a free parameter of the geometry to be minimized for the particular collider event in comparison, comparable to the thrust axis.}
\end{figure}

Unlike event isotropy, which exhibits continuous rotational symmetry, the two particle configuration is not a single event, but a manifold of events related by global azimuthal rotations.
This set of reference geometries, \pbb, is defined as
\begin{equation}
\mathcal{M}_2(\theta) = \left\{\frac{1}{2}\left( \cos\theta, \sin\theta \right), -\frac{1}{2}\left( \cos\theta, \sin\theta \right) \right\}
\end{equation}
where $\theta$ is the relative angle of the event, and the factors of $1/2$ are from normalizing each particle energy by the total energy.
One additional aspect of determining the dipole distance \evCol is to minimize the EMD over the orientation of the reference geometry with respect to the collider event. 
This is analogous---but not identical---to finding the thrust axis, or the axis along which the event is most collimated.
In practice, the minimization over this angle is  performed numerically using standard libraries when computing the manifold distance, \emph{e.g.} Python Optimal Transport~\cite{flamary2017pot}, which is used throughout this result.


\subsection{Energy Weight}
As we consider only transverse information, the radiation pattern of a set of massless particles is fully specified as a set of objects with two coordinates: scalar transverse momentum \pt~and azimuthal angle $\phi$.
The transverse momentum of each particle serves as the energy weight to be moved by transport function $f_{ij}$ defined in \Eq{eq:EMDdef}. 
Since the total energy of the event is irrelevant for event-shape observables, we normalize the transverse momentum of each particle by the scalar sum 
\begin{equation}
p_{Ti} \rightarrow \frac{p_{Ti}}{\sum p_{Ti}} \equiv E_i
\label{eq:engNorm}
\end{equation}
such that the total (dimensionless) energy is $E_\text{tot} =1$; \emph{n.b.} that the reference geometries defined in \Sec{sec:RefGeo} also are normalized to unit energy by construction.
The transport function $f_{ij}$ therefore specifies what fraction of the total energy in an event is moved from position $i$ in the collider event to position $j$ in the reference geometry.

\subsection{Distance Measure}
\begin{figure}[t!]
    \centering
    \includegraphics[width=0.7\textwidth]{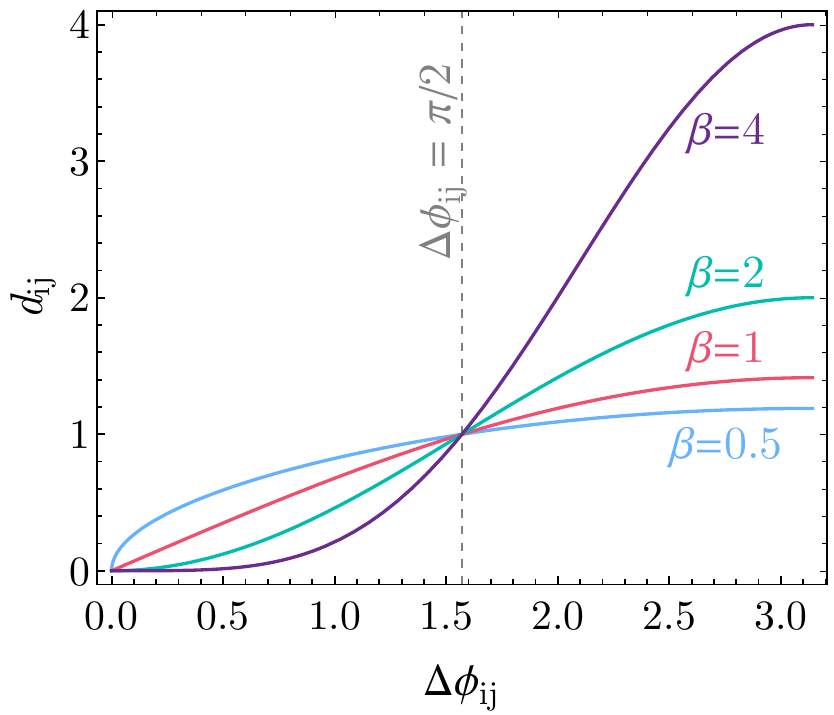}
    \caption{The distance weight as a function of the angular separation for $\beta = 0.5, 1, 2, 4$. We see that distances across small angular separation ($\Delta \phi_{ij} < \pi/2$) are penalized more strongly for low $\beta$, whereas large angular separation  ($\Delta \phi_{ij} > \pi/2$) is much more heavily penalized for higher $\beta$.}
    \label{fig:distComp}
\end{figure}

The distance measure is a function of only the angular distance between two particles 
\begin{equation}
    d_{ij} = D(\Delta\phi_{ij}).
\end{equation} 
While any monotonic function can be used, in this work we focus on measures of the form 
\begin{equation}
    d_{ij} = \left( 1 - \cos\Delta\phi_{ij}\right)^{\beta/2}
\label{eq:distDef}
\end{equation}
for $\beta = \{0.5, 1, 2, 4\}$. 
In the small angle limit, $\beta=0.5$ corresponds to the Les Houches Angularity (LHA) \cite{Andersen:2016qtm}, $\beta = 1$ is similar to width or jet broadening \cite{Andersen:2016qtm, Catani:1992jc,Rakow:1981qn, Ellis:1986ig}, $\beta = 2$ is similar to thrust \cite{Farhi:1977sg}, and $\beta =4$ is used to further exaggerate observed trends. 
A visualization of these distance weights is shown in \Fig{fig:distComp}, as a function of the angular separation between particles. 
We choose this distance function and set of parameters to construct a set of effective benchmarks with which to understand the efficacy of these observables. 
By changing $\beta$, the user can control how strongly small angled splittings are penalized relative to large angled splittings. 
This is clearly illustrated in \Fig{fig:distComp}: smaller values of $\beta$ lead to a relatively higher cost for moving particles with small angular splittings $(\Delta \phi_{ij} < \pi/2)$, where as for larger values of $\beta$ it becomes essentially free to move particles small distances.
This is particularly relevant in the context of moving particles through distances comparable to the jet radius used in experimental studies: for larger values of $\beta$, movement of soft particles at the scale of a jet is essentially free.
In the same vein, the cost of moving particles across larger distances $(\Delta \phi_{ij} > \pi/2)$ becomes much more expensive for larger values of $\beta$. 

The most effective or discriminating choice of $\beta$ is  dependent on the underlying physics of the analysis. 
In the remainder of this work, we will show various examples that illustrate how difference choices of reference geometries and distance metrics can impact the phenomenology of the manifold distance.


\subsection{Analytics}
In this section we review some of the relevant analytics when computing and understanding the various manifold distances.
As detailed discussions on how to compute or approximate event isotropy for a given $\beta$ are already presented in Refs.~\cite{Cesarotti:2020hwb, Cesarotti:2020ngq, Cesarotti:2020uod}, we do not explicitly review how to analytically compute this observable. 
Instead, we present illustrative examples of the EMD computed with simple radiation patterns to better understand the range of these observables.

\begin{figure}[htbp]
    \centering
    \subfigure[]{
    \includegraphics[width=0.45\linewidth]{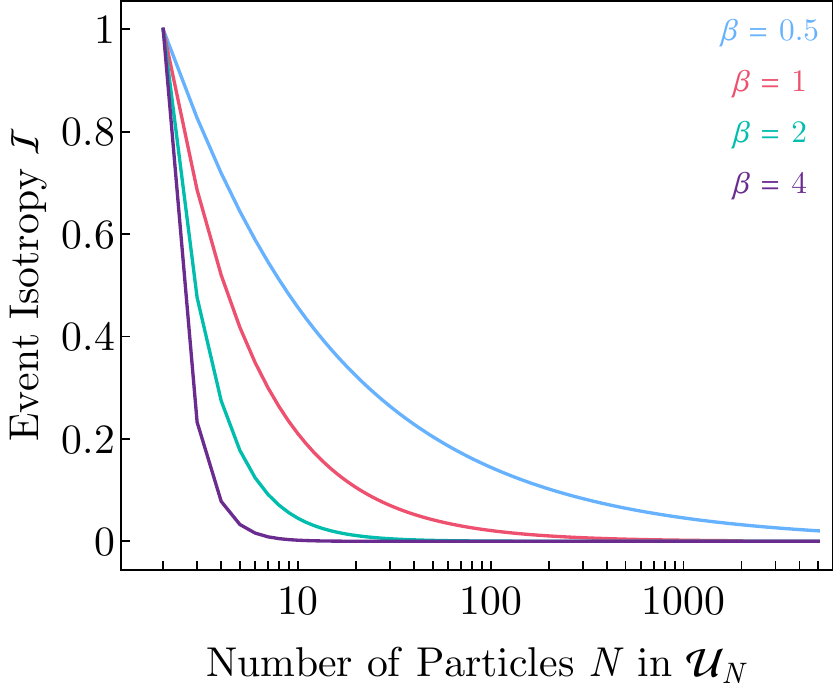}} \qquad
    \subfigure[]{
    \includegraphics[width=0.47\linewidth]{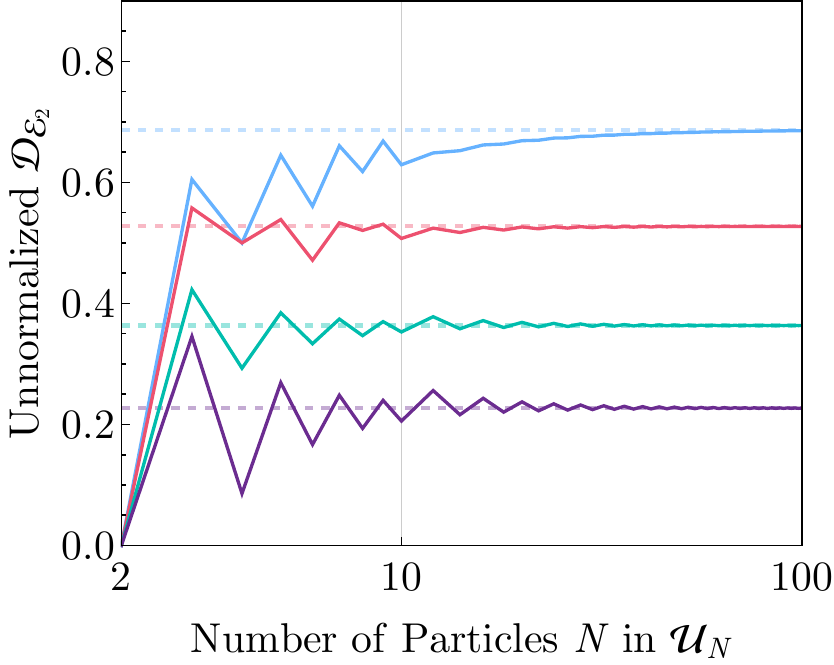}}
    \caption{Manifold distances as a function of $N$-particle symmetric events (meaning equal energy, evenly distributed particles). 
    The values of event isotropy are shown on the left and exhibit a monotonic behavior, all maximized for the 2-particle (\pbb) configuration, decreasing as the $N$-particle event approaches the continuous limit.
    The values of (unnormalized) dipole distance \evCol, shown on the right, are not monotonic, and jump around before asymptoting to the infinite particle limit.
    We show the unnormalized dipole distance values simply for ease of viewing.
    The dashed lines indicate the value of dipole distance computed for $N = \infty$.
    To the right of the solid line we only plot the value for $N$ divisible by 2 for ease of analytic computation.
    }
    \label{fig:normalize}
\end{figure}
The range of the manifold distances is closed and well defined for events with sets of massless particles and zero net momentum (`recoil-corrected,' following the discussion in Ref.~\cite{Banfi:2010xy}).
Therefore, for ease of interpretation, subtract any unbalanced momentum from the event and normalize observables to their maximal value such that their domain is between 0 to 1. 
However, it is not immediately obvious what the maximally distant radiation pattern is. 
To show this, we plot the manifold distances as a function of symmetric $N$-particle radiation patterns $\mathcal{U}_N$ in Fig.~\ref{fig:normalize}

Since the reference geometry for event isotropy is continuous and symmetric under rotations, an analytic solution exists for the distance computed to events with discrete rotational symmetry. 
If an event has $N$ uniformly spaced particles of equal weight $1/N$, the manifold distance is the work it takes to spread the particle evenly over the distance spacing:
\begin{equation}
\text{EMD}(\mathcal{U}_N, \mathcal{U}_\infty) = N \times \int_{-\pi/N}^{\pi/N} \frac{d \phi}{2\pi} (1-\cos\phi)^{\beta/2}
\label{eq:evIsoNorm}
\end{equation}
Note that while other isotropic manifold distances can have closed forms, the symmetry is necessary for the calculation done in Eq.~\ref{eq:evIsoNorm}.
More details on analytic calculations can be found in Ref.~\cite{Cesarotti:2020hwb}.

For event isotropy, the values are saturated at any \pbb~event for all values of $\beta$ and monotonically decrease as $N$ approaches infinity.
It is notable that for smaller values of $\beta$, the approach to 0 is much more gradual, which we see in Fig.~\ref{fig:normalize}a.
Because small $\beta$ weights smaller angular displacements relatively more than larger $\beta$ values, event isotropy computed with small $\beta$ is more sensitive to the discrete nature of the event.
Another interpretation of this plot is that the curves bound the smallest value of event isotropy that can be achieved by an $n$-particle event.

The computation is less straightforward for dipole distance. 
For distance measures of $\beta \geq 1$, the maximally-distant configuration to the back-to-back reference geometry is the three-prong event. 
However, for $\beta = 0.5$, we see in Fig.~\ref{fig:normalize}b that the maximally distant configuration from the dipole event is the uniform radiation pattern. 

One more added complexity to computing dipole distance is the minimization over angular configuration. 
For simple events, such as the discretized uniform event, symmetry arguments reduce the possible configurations that could be minimal (effectively either aligned with a particle or aligned along a midpoint between particles), and analytic calculations are still tractable. 
However, for more complicated events, it is often easiest (at the expense of calculation cost) to minimize the orientation of \pbb~numerically.

To normalize the range of the observable from 0 to 1, we divide the manifold distance by the maximal distance for that geometry and distance metric. 
In order to make results  more readily interpretable, we plot \evCol~and \evIso, where
\begin{equation}
    \bar{\mathcal{I}} \equiv 1 - \mathcal{I}
\end{equation}
such that observable values near 0 correspond to collimated events, and isotropic radiation patterns have values near 1.

\section{Simulation Results}
\label{sec:results}

\subsection{Monte Carlo setups and selections}\label{sec:selection}

To demonstrate the behaviour of manifold distances constructed with various distance metrics and reference geometries, we compute the spectrum of the observables for two Monte Carlo (MC) samples. 
The simulations were produced with \texttt{Pythia} v8.306 at a center-of-mass energy $\sqrt{s}=13$~TeV, using the default `Monash' tune~\cite{Skands:2014pea}.
No detector simulation is applied, although selection cuts made to objects (outlined below) are intended to resemble realistic experimental selection criteria.
Similarly, no studies of the impact of pileup are performed.
These effects were found to be negligible in the measurement performed by ATLAS using high-\pt{}~jets~\cite{STDM-2020-20}, and in the measurement performed by CMS using charged hadrons~\cite{CMS:2024bwe}.
Two samples of events were generated, which provide samples of events with distinct characteristics that allow the behaviour of manifold distances to be illustrated.

The first sample consists of groups generated using the default \texttt{Pythia} group of hard QCD $2\rightarrow2$ processes, which we call the `Hard QCD' (`HQCD') sample.
The implementation of the anti-$k_t$ algorithm with radius $R=0.4$~\cite{Cacciari:2008gp} from \texttt{FastJet} is used~\cite{Fastjet,Cacciari:2005hq} to reconstruct jets from stable final-state particles with a transverse momentum above $500$~MeV and within $|y|<4.9$, excluding neutrinos.
Such a requirement approximates the acceptance of detectors such as ATLAS or CMS at the LHC.
The events in this sample are required to have at least two anti-$k_t$~jets with a \pt{}~above $60$~GeV and rapidity $|y|<4.5$~to be included in the study.
Additional jets are also retained if they are present and satisfy the same criteria.

The second sample consists of events generated using the default \texttt{Pythia} group of soft QCD processes, which we call the `Soft QCD' (`SQCD') sample. 
This sample resembles `minimum bias' events at the LHC.
Jets are not typically present in such events, and so the only criteria is that at least two particles with \pt{}~above 500~MeV are present within $|y|<4.9$, excluding neutrinos.

Since these two samples have notably different global behavior, they are illustrative to how the two reference geometries (as well as their distance metrics) can be sensitive to different features of the radiation patterns.
For both samples, the manifold distances are primarily computed using the four-momenta of all hadrons in the event with \pt{} above 500~MeV and $|y|<4.9$, but we also consider using jets with \pt{}~above 60~GeV and $|y|<4.5$ as the input objects for some comparisons using the HQCD sample.

\subsection{1d Distributions}
\begin{figure}[htbp]
\centering
\subfigure[]{\includegraphics[width=0.45\textwidth]
{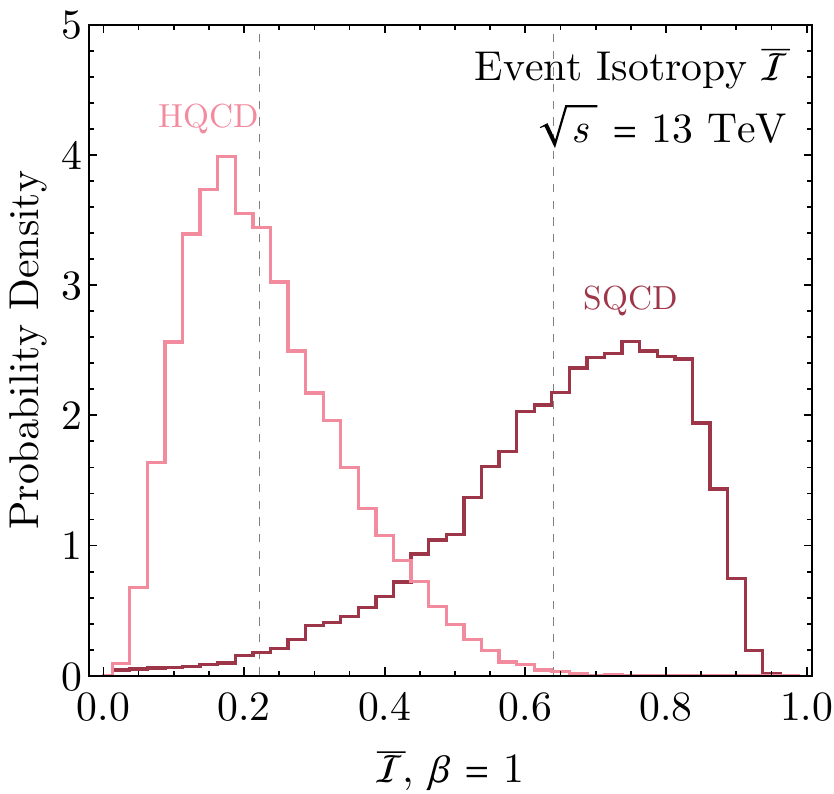}} \quad
\subfigure[]{\includegraphics[width=0.45\textwidth]
{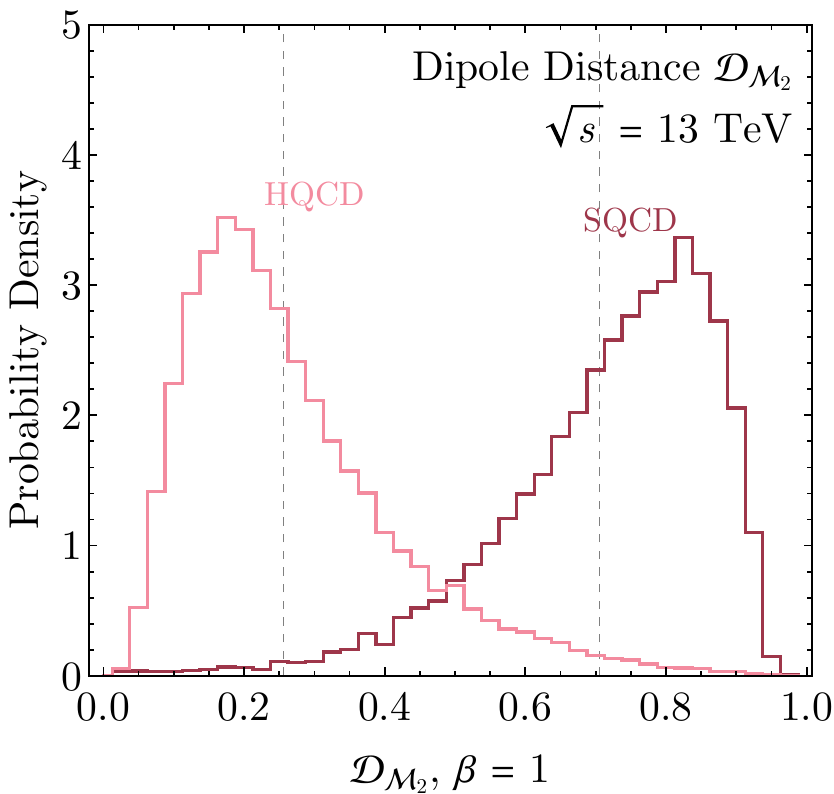}}
\caption{Distributions in event isotropy \evIso~on the left and dipole distance \evCol~on the right for both hard and soft QCD samples. The dashed lines indicate the means of the distributions.}
\label{fig:1dbeta1}
\end{figure}

To begin illustrating how these observables behave for different physical processes, we first consider the 1d distributions for the various reference geometries, values of $\beta$, and samples of simulated events.
We expect that for the HQCD sample, the distributions will be peaked more towards $ \sim 0$ (back-to-back), whereas the SQCD will be pushed towards $\sim 1$ (isotropic). 
To illustrate the difference for the different samples, we compare the distributions for both manifold distances using $\beta = 1$ in Fig.~\ref{fig:1dbeta1}. 

In this figure, we note a few important features of the distributions. 
As expected, both observables demonstrate that the SQCD sample is more isotropic than the HQCD sample. 
As we have stated previously, manifold distances (and similar observables like thrust) have a greater dynamic range when the distance is computed with respect to a similar reference geometry than a very distant one: being `far' from a radiation pattern is often a more degenerate mapping into a manifold distance.
This illustrates how important the choice of reference geometry is for signal discrimination: a signal well-described by the underlying geometry will have support across the observable's range, while a background poorly matched by the reference geometry will be pushed towards the extremal values, as indicated by the dashed lines at the mean of the distribution in Fig.~\ref{fig:1dbeta1}.
This feature reinforces the notion that `further' events tend to saturate more quickly than `nearby' events.

\begin{figure}[htbp]
\centering
\subfigure[]{\includegraphics[width=0.45\textwidth]
{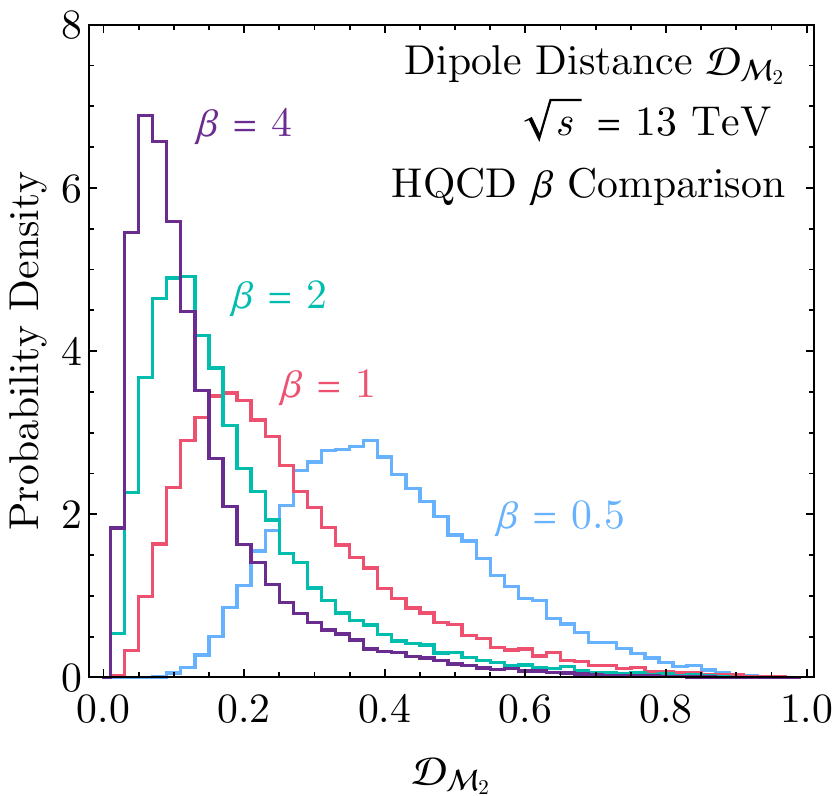}} \quad
\subfigure[]{\includegraphics[width=0.45\textwidth]
{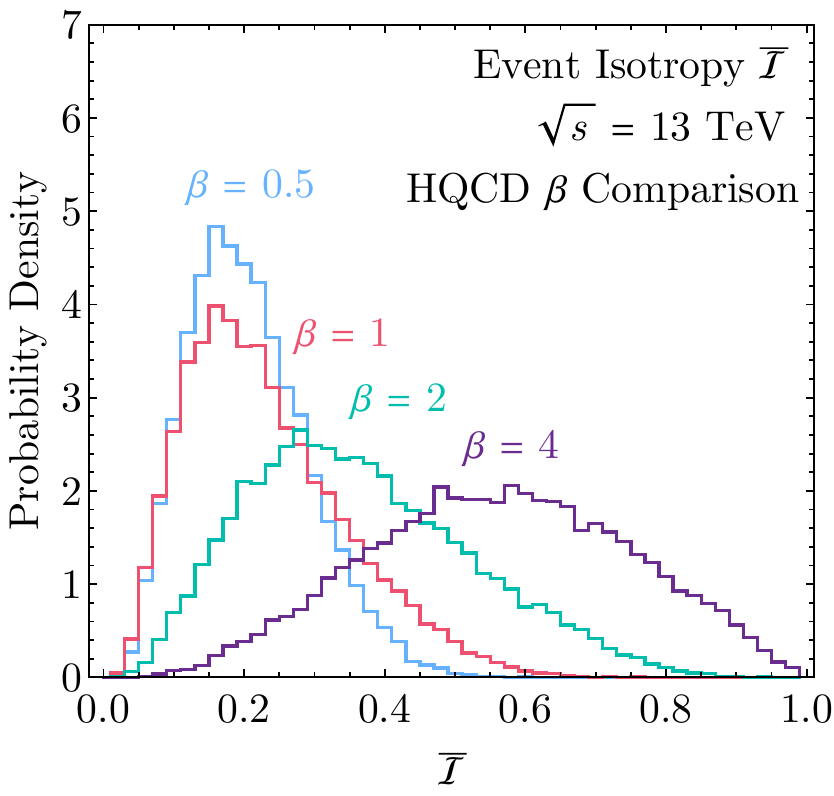}}
\subfigure[]{\includegraphics[width=0.45\textwidth]
{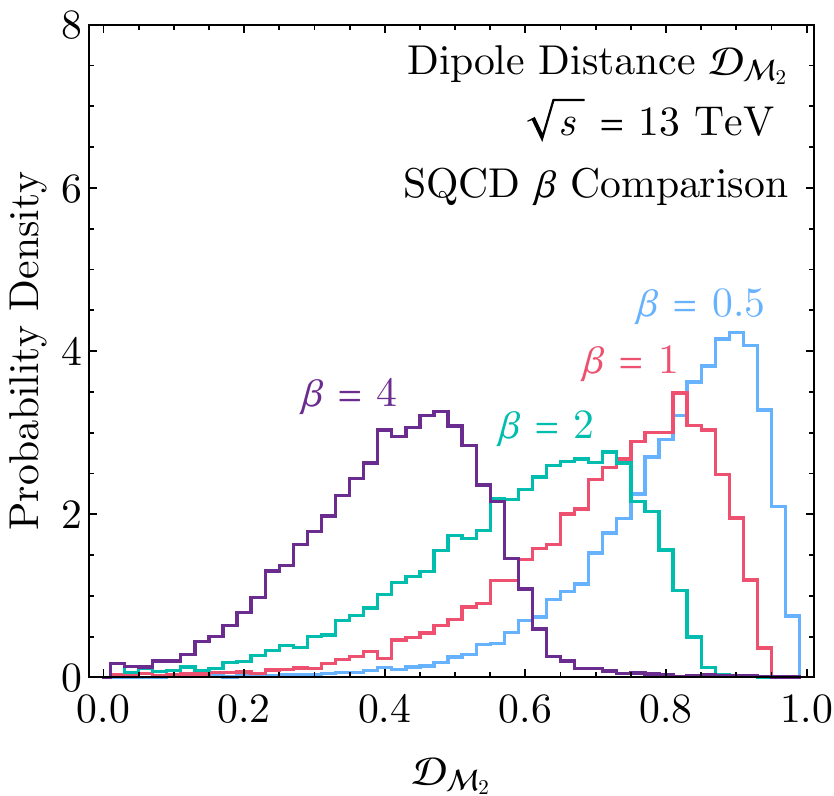}} \quad
\subfigure[]{\includegraphics[width=0.46\textwidth]
{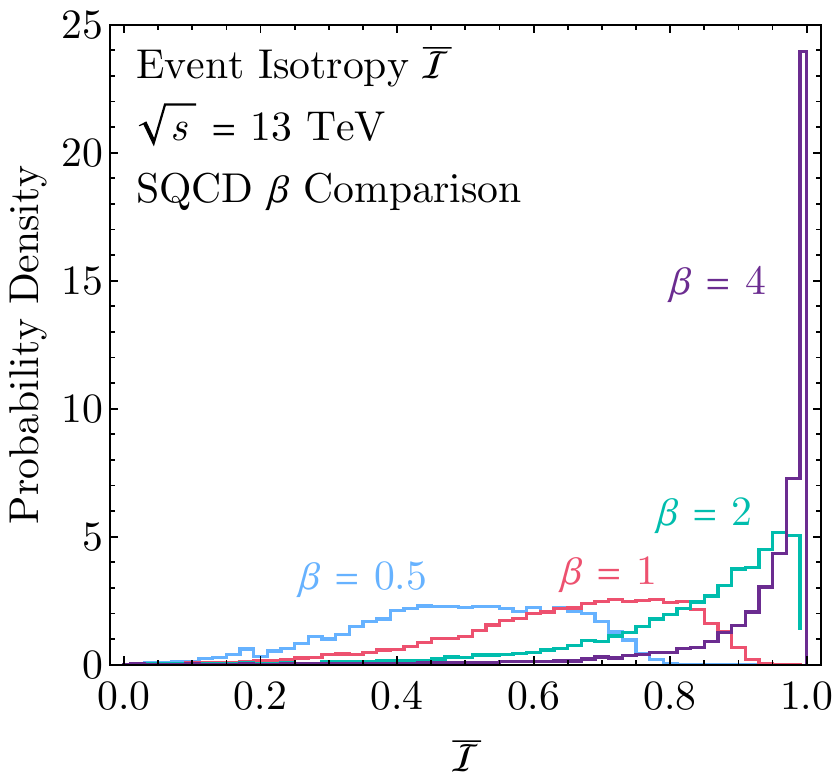}}
\caption{In each plot the distributions for the HQCD (top) and SQCD (bottom) samples are shown for all choices of $\beta$. The plots on the left show the \evCol~and those on the right are \evIso. We see for both samples, the distributions are more strongly peaked when computed with their `more correct' reference geometry, but the ordering of the means of the distributions are inverted between the two geometries.}
\label{fig:allBeta1d}
\end{figure}

To understand how different choices of $\beta$ can affect these observables, we plot all the distributions for both reference geometries, both samples, and all four values of $\beta$ in Fig.~\ref{fig:allBeta1d}. 
We see that the effects of changing $\beta$ can be as significant as our observations regarding the choice of reference geometry. 

Let's first compare what is similar between the HQCD and SQCD samples. 
From Fig.~\ref{fig:allBeta1d}, we see that for dipole distance, higher $\beta$ values push the distribution towards 0, meaning the events look `jettier,' on average.
However; for event isotropy, larger $\beta$ values push the distribution towards 1, making events look more isotropic, on average. 
We can explain this behavior by revisiting Fig.~\ref{fig:distComp}, where we see that as $\beta$ is increased, the cost of moving particles that are nearby falls rapidly. 
Thus, as the cost to move particles already somewhat close to the reference geometry goes down, the values skew toward being closer to the reference geometry. 

We note that while the distributions are most strongly peaked for the `correct geometry' (dipole for HQCD, and quasi-isotropic for SQCD) for the largest value of $\beta$, the most strongly peaked for the `opposite geometry' is the \emph{smallest} value of $\beta$.
To understand why, we must remind ourselves of the full logic of how $\beta$ affects the distributions: larger $\beta$ weights small movements \textit{less} and large movements \textit{more}, whereas small $\beta$ weights small and large movements nearly equally.
Hence, for smaller values of $\beta$, the distributions shift away from the extremal values as the EMD calculation is more sensitive to small perturbations away from the reference geometry.

From these plots we can make some generic statements about the observables. 
\begin{itemize}
\item If one expects the reference geometry is a good approximation of their collider event set, using \textit{smaller} values of $\beta$ provides a larger dynamic range and can be used to discriminate similarly `shaped' underlying physics. 
\item If the reference geometry is far from the collider event set, a \textit{larger} value of $\beta$ can provide a greater dynamic range as the increased distances between particles are penalized more strongly. 
\end{itemize}
While we include $\beta=4$ for the sake of study, the fact that we already see saturation at perfectly isotropic for event isotropy in the soft QCD sample suggests that using large $\beta$s can identify if there is approximate underlying symmetries in the set of events.

\subsection{2d Correlations}
Now that we have understood the difference in the behaviors of the observables for various $\beta$ values and samples, let's consider the degree to which the manifold distances are correlated. 
We will look at the correlations between the observables both for different values of $\beta$ while fixing the reference geometry, and fixing $\beta$ while comparing the reference geometry.

\subsubsection*{Correlations between $\beta$}
Let's begin by considering the correlations between the same observables (\emph{i.e.} same reference geometries) for various values of $\beta$.
For ease of plotting we do not consider every combination of $\beta$; we consider enough to be indicative of any global trends. 

\begin{figure}[htbp]
\centering
\subfigure[]{\includegraphics[width=0.3\textwidth]
{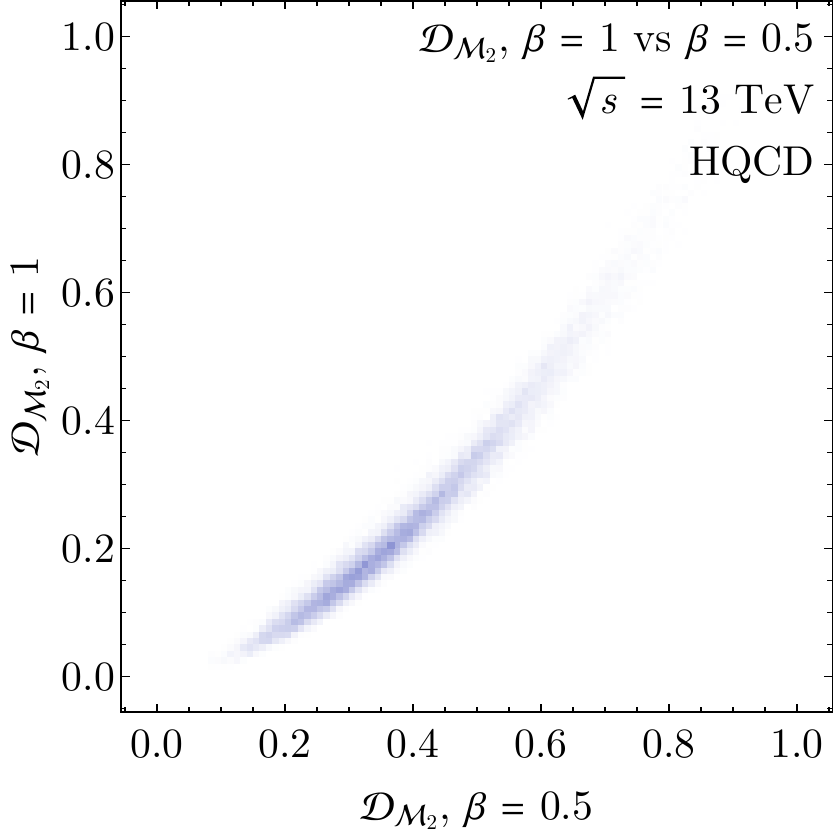}} \quad
\subfigure[]{\includegraphics[width=0.3\textwidth]
{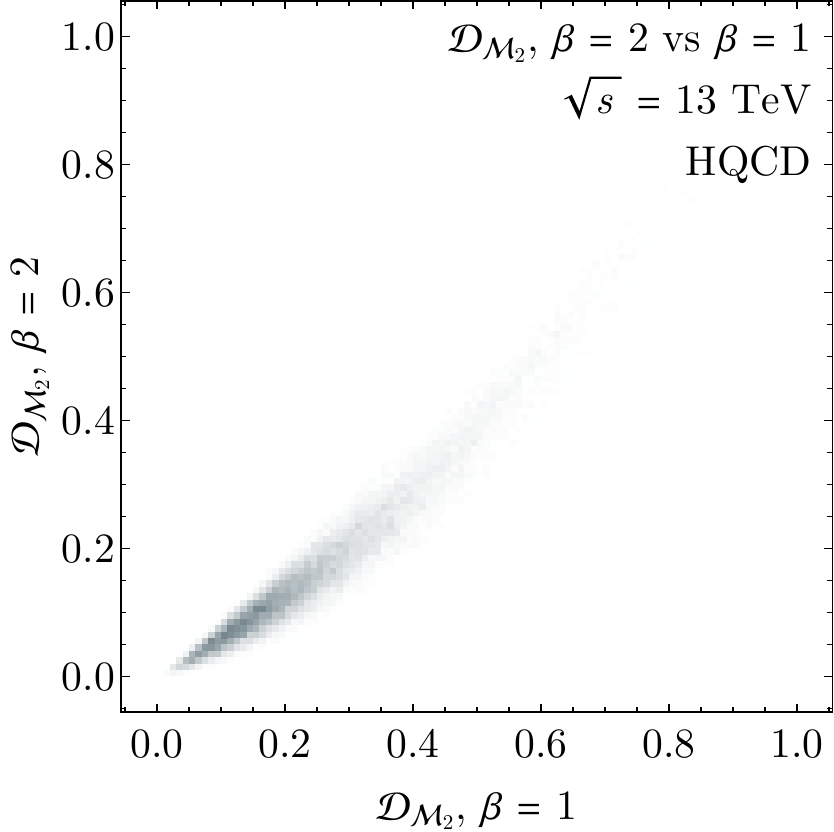}} \quad
\subfigure[]{\includegraphics[width=0.3\textwidth]
{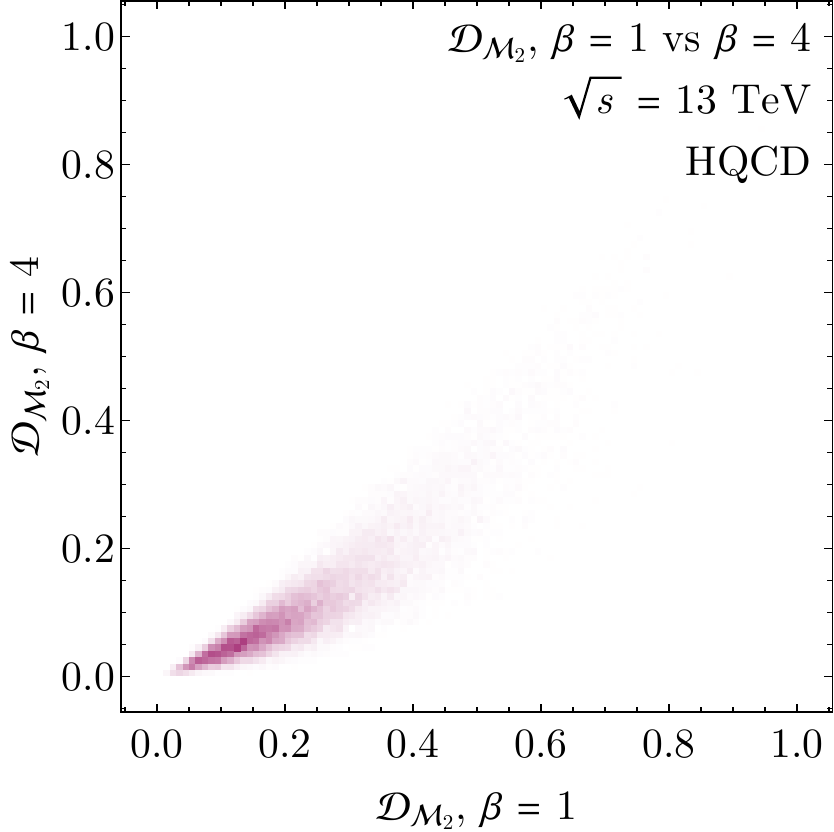}}
\subfigure[]{\includegraphics[width=0.3\textwidth]
{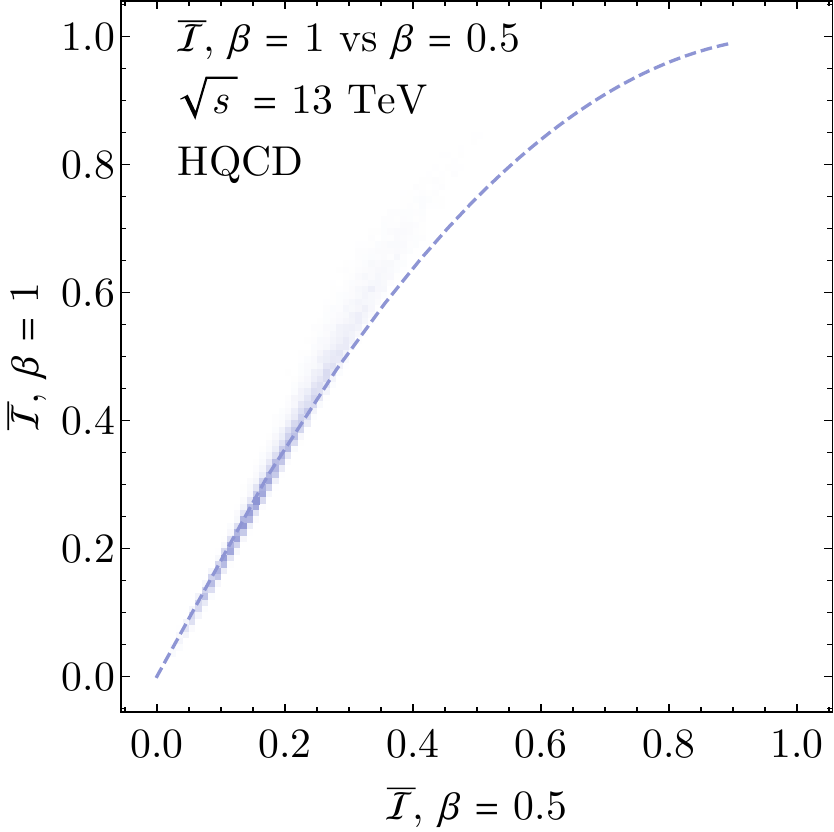}} \quad
\subfigure[]{\includegraphics[width=0.3\textwidth]
{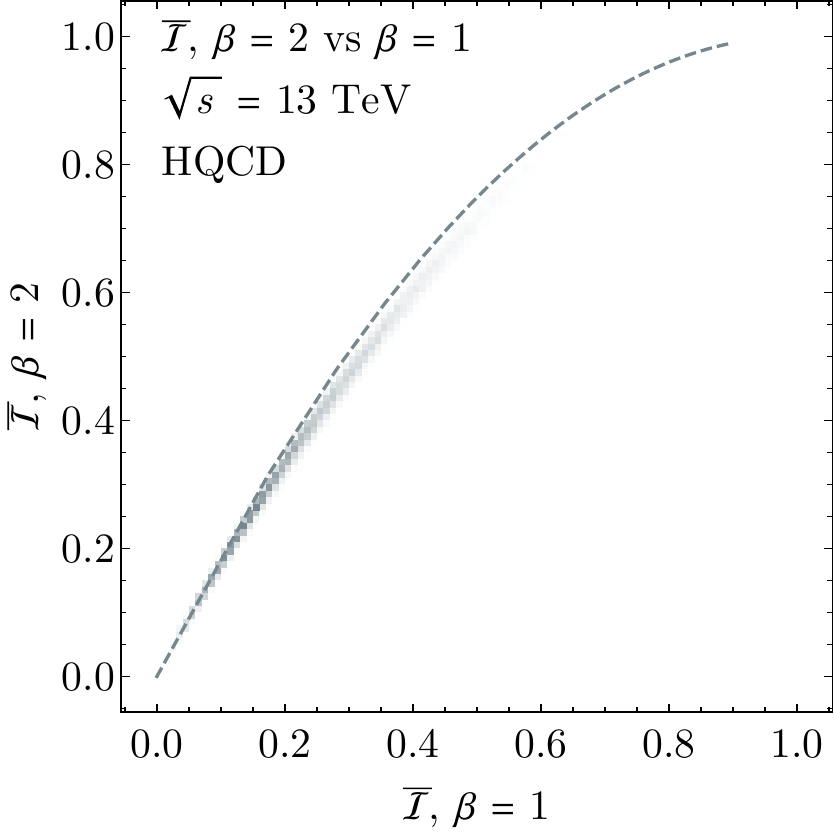}} \quad
\subfigure[]{\includegraphics[width=0.3\textwidth]
{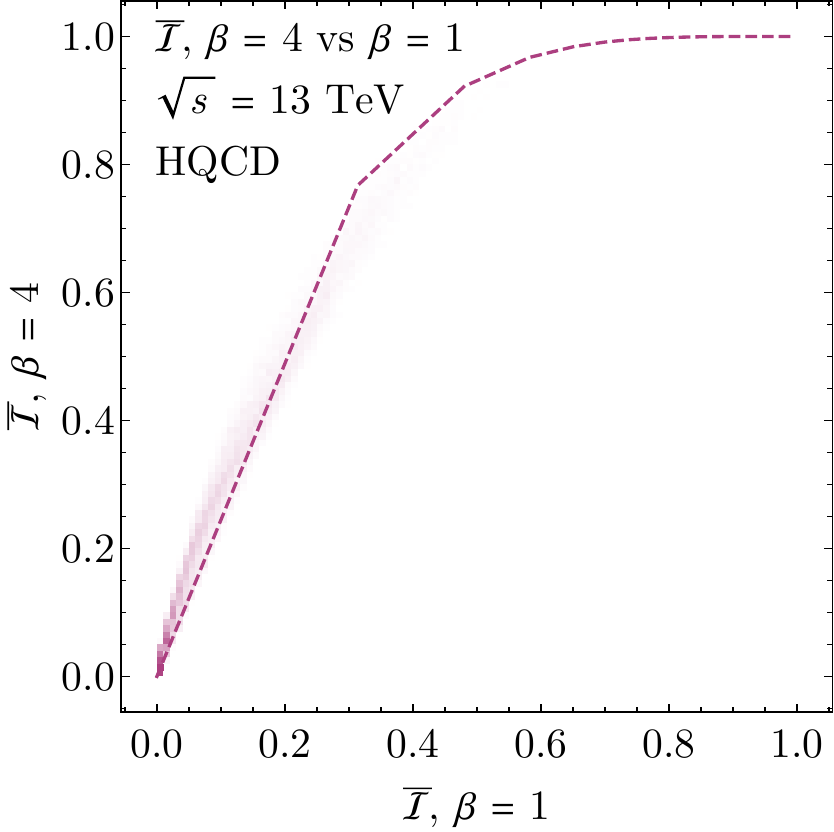}}
\caption{Plots of the 2d distributions of \evCol~(top) and \evIso~(bottom) for the HQCD sample.
As a simple analytic model of the underlying behavior, we plot the curve for the maximally symmetric finite-multiplicity event (\evIso~ of $\mathcal{U}_N$) as a dashed line; this line represents the exact correlation we would expect if the samples were maximally symmetric.
We see that \evIso~is always more correlated, and that the correlations are less between high $\beta$ and low $\beta$.}
\label{fig:2dHQCD}
\end{figure}

We start by looking at the HQCD samples in Fig.~\ref{fig:2dHQCD}. 
It is immediately evident that for all values of $\beta$, the distributions are much more correlated for event isotropy than for dipole distance.
We conjecture that this is due to the large underlying symmetry of the reference geometry for event isotropy. 
On average the distances traveled by particles to get to the reference geometry of a dipole rather than the isotropic distribution are greater, and the non-linearities in the EMD calculation are more evident.

For the HQCD sample in particular, the disparity in correlations for the different observables is also due to the fact that the sample is better approximated as a dipole than an isotropic event. 
Because manifold distances are most powerful at discriminating `close' events than `far' events (see Ref.~\cite{Cesarotti:2020hwb}), the event isotropy distributions for HQCD tend to saturate the anisotropic regime and appear more correlated.

We also note the general trend that when the values of $\beta$ are more different, the observables become less directly correlated (\emph{i.e.} there is more spread in the 2d distribution).
This is a statement underscoring how different choices of $\beta$ fundamentally measure different features of the radiation pattern: high $\beta$ is insensitive to small movements, whereas smaller $\beta$ weights small and large distances increasingly equally. 
The increased spread demonstrates that, depending on the analysis at hand, studying two or more values of $\beta$ might lead to improved performance.

\begin{figure}[htbp]
\centering
\subfigure[]{\includegraphics[width=0.3\textwidth]
{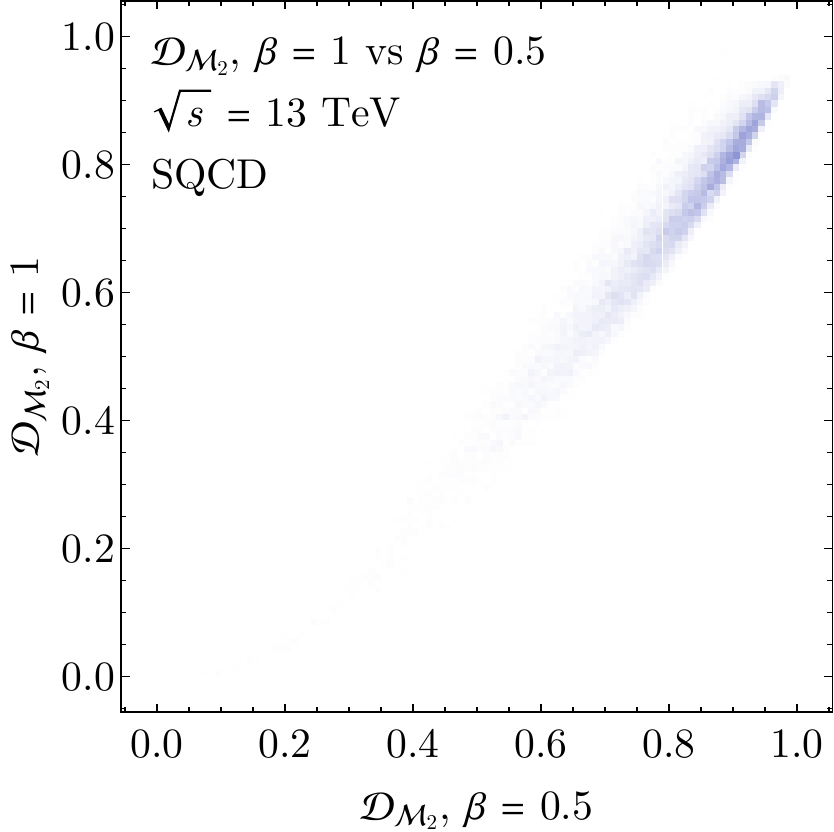}} \quad
\subfigure[]{\includegraphics[width=0.3\textwidth]
{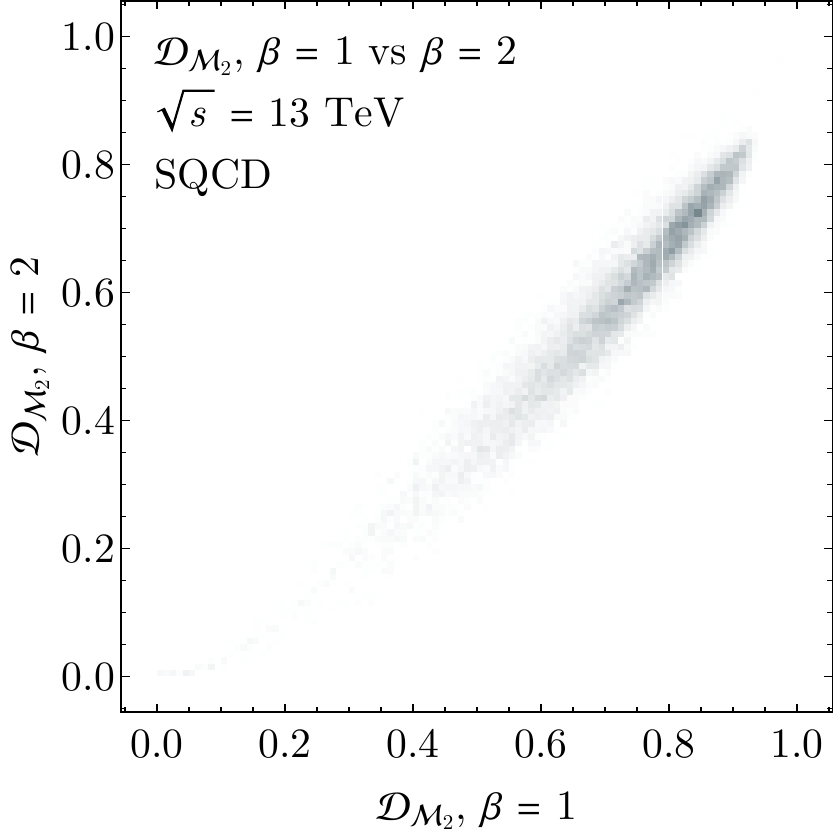}} \quad
\subfigure[]{\includegraphics[width=0.3\textwidth]
{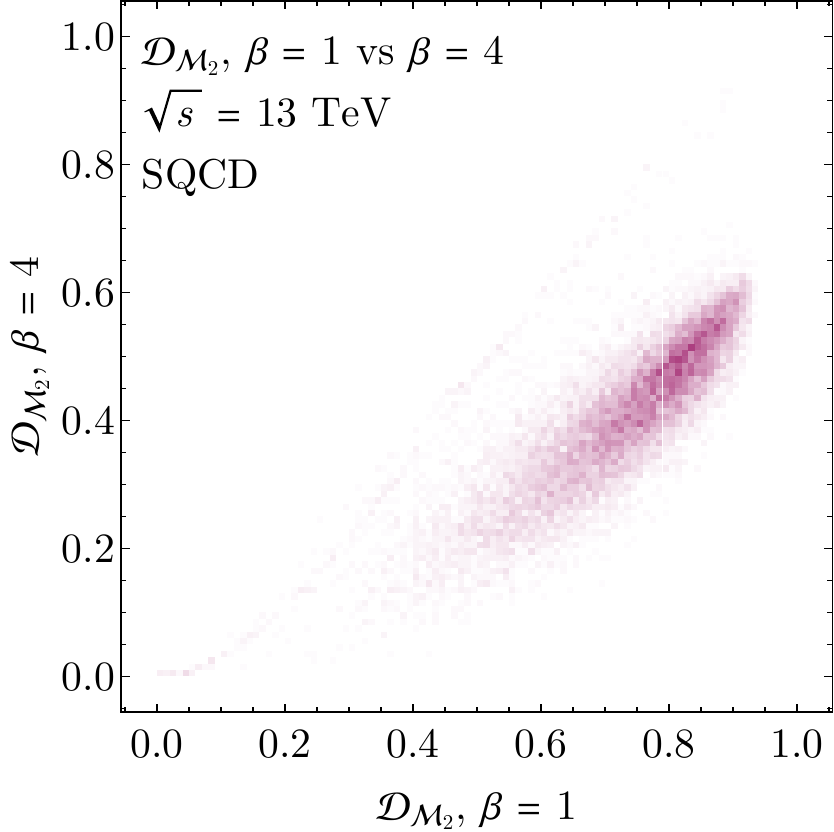}}
\subfigure[]{\includegraphics[width=0.3\textwidth]
{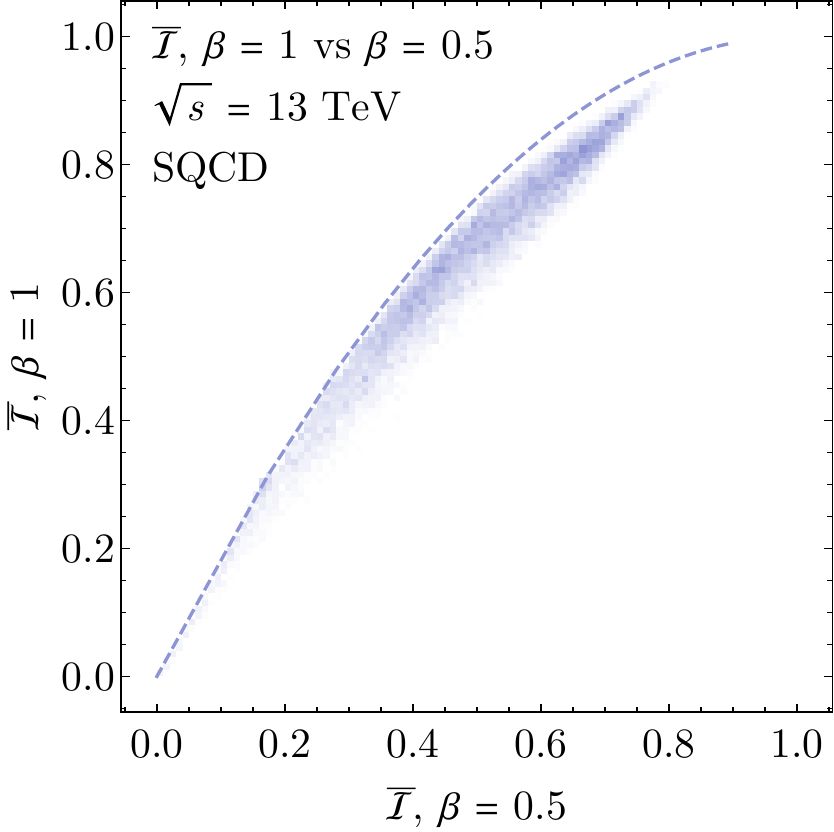}} \quad
\subfigure[]{\includegraphics[width=0.3\textwidth]
{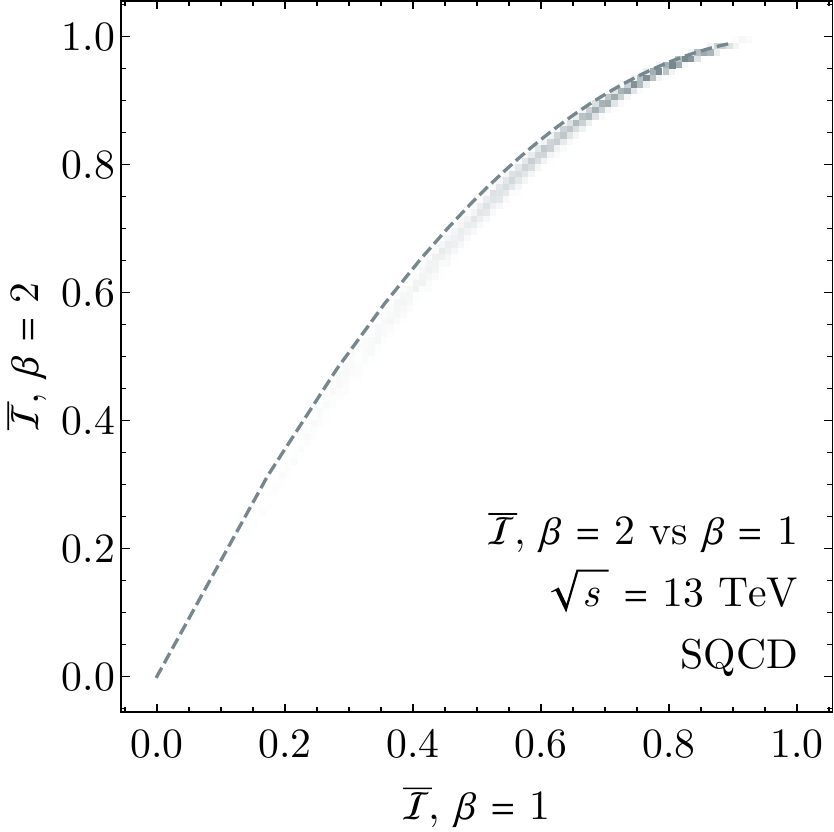}} \quad
\subfigure[]{\includegraphics[width=0.3\textwidth]
{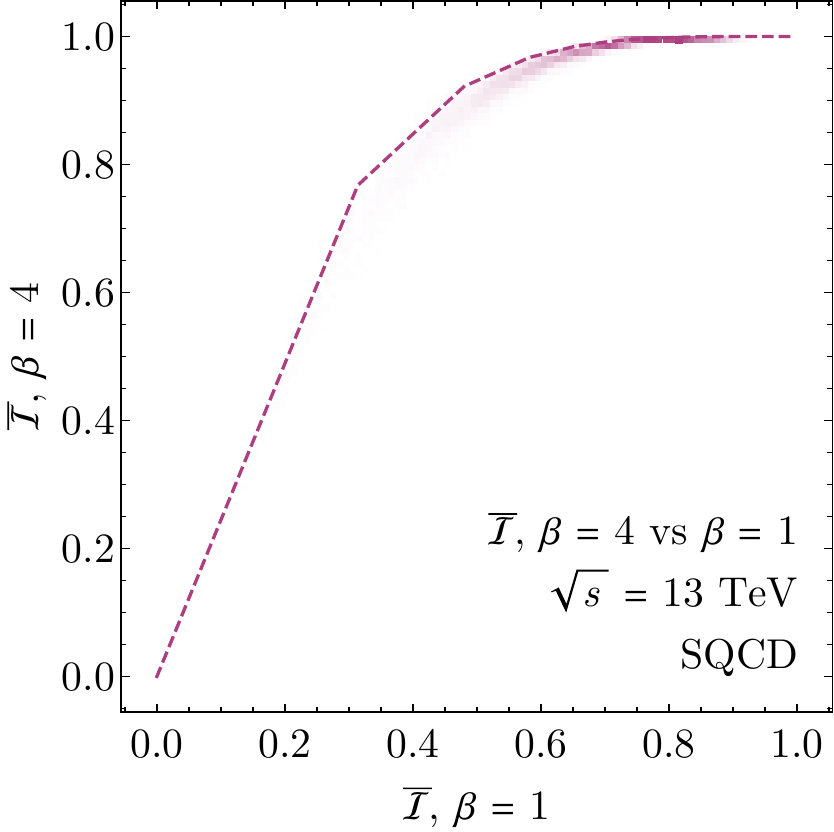}}
\caption{The same figure as Fig.~\ref{fig:2dHQCD}, but with the SCQD sample. We see similar trends, but \evIso~is less correlated for SQCD than HQCD. Note that in the SQCD sample has a feature where the boundary to the left of the population looks defined. These are events where only two particles survived the kinematic cuts, and should not be taken as representative of the sample.}
\label{fig:2dsoft}
\end{figure}
Now, let's move our attention to the SQCD samples to confirm our statements are generic to the observables, and not specific to dijet-like radiation patterns: the same plots are shown for SQCD in Fig.~\ref{fig:2dsoft}. 
Again, we generally find that dipole distance is less correlated than event isotropy, except for the case of $\beta=0.5$ and $\beta=1$. 
The argument for this behavior is the same as that for the HQCD samples, but with the difference that for sufficiently small $\beta$, even little distances become relevant in the event isotropy calculation when the collider event is also somewhat isotropic. 
As we increase $\beta$, the distance weight for nearby particles becomes so small that the distributions become highly correlated. 
From the plots in Fig.~\ref{fig:2dsoft}, we see that the correlation collapses to the curve for the maximally isotropic value for a fixed number of particles. 

Overall, we can conclude that the choice of $\beta$ when computing event isotropy doesn't change the analysis as significantly as the choice of $\beta$ for dipole distance. 
The biggest exception we see to this statement is small $\beta$ for quasi-isotropic samples can produce different results as it is more sensitive to small perturbations from the reference geometry.
Therefore, we recommend that unless there is a particular motivation to use a specific $\beta$, one should use whichever is least computationally expensive for event isotropy\footnote{\emph{i.e.} avoiding square roots and other expensive operations when possible.}. 
However, the choice of $\beta$, especially large vs. small, can noticeably change the distribution of dipole distance.
The choice of $\beta$ should then be explored to optimize the analysis at hand.

\subsubsection*{Correlations between manifold distances}
\begin{figure}[htbp]
\centering
\subfigure[]{\includegraphics[width=0.22\textwidth]
{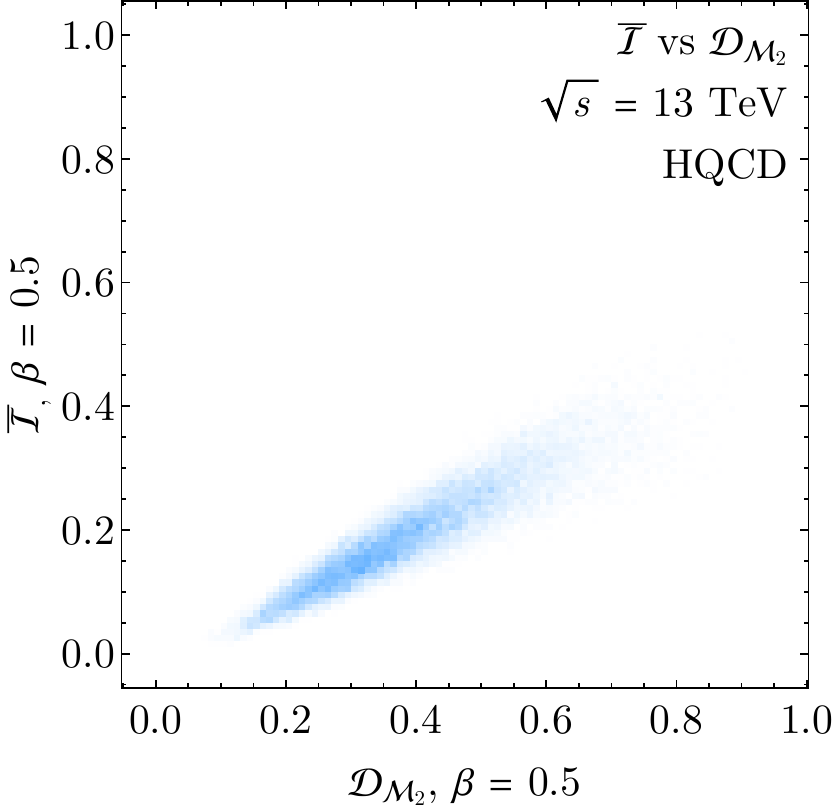}} \quad
\subfigure[]{\includegraphics[width=0.22\textwidth]
{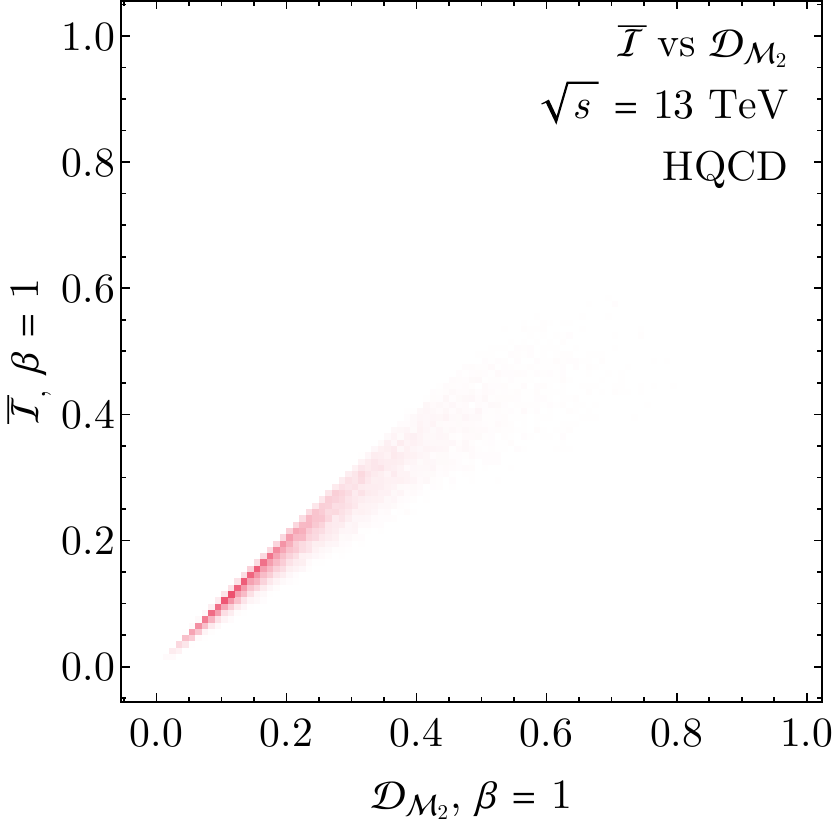}} \quad
\subfigure[]{\includegraphics[width=0.22\textwidth]
{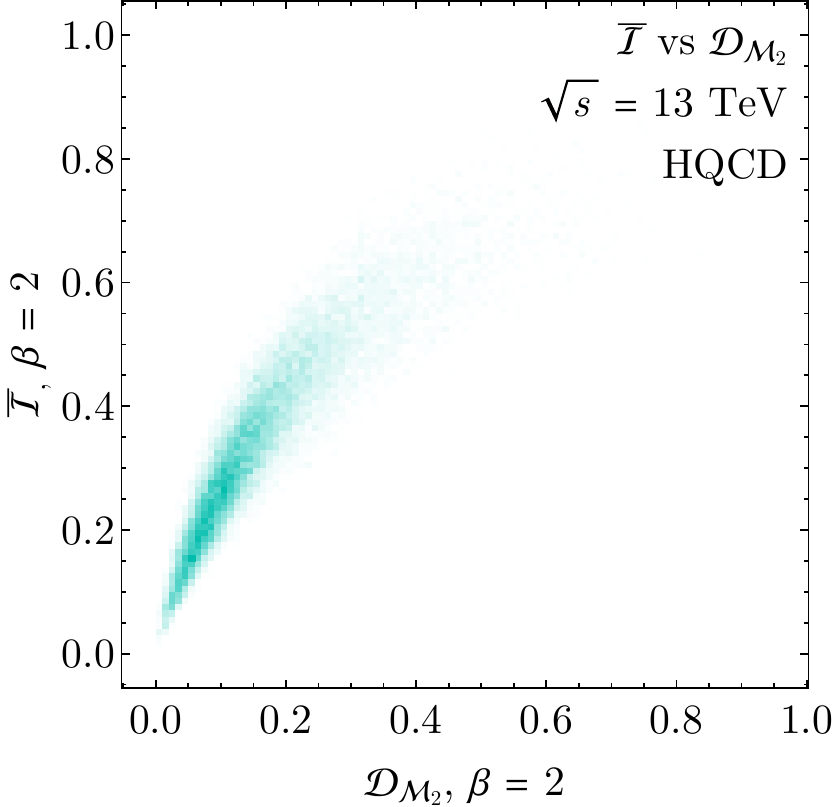}} \quad
\subfigure[]{\includegraphics[width=0.22\textwidth]
{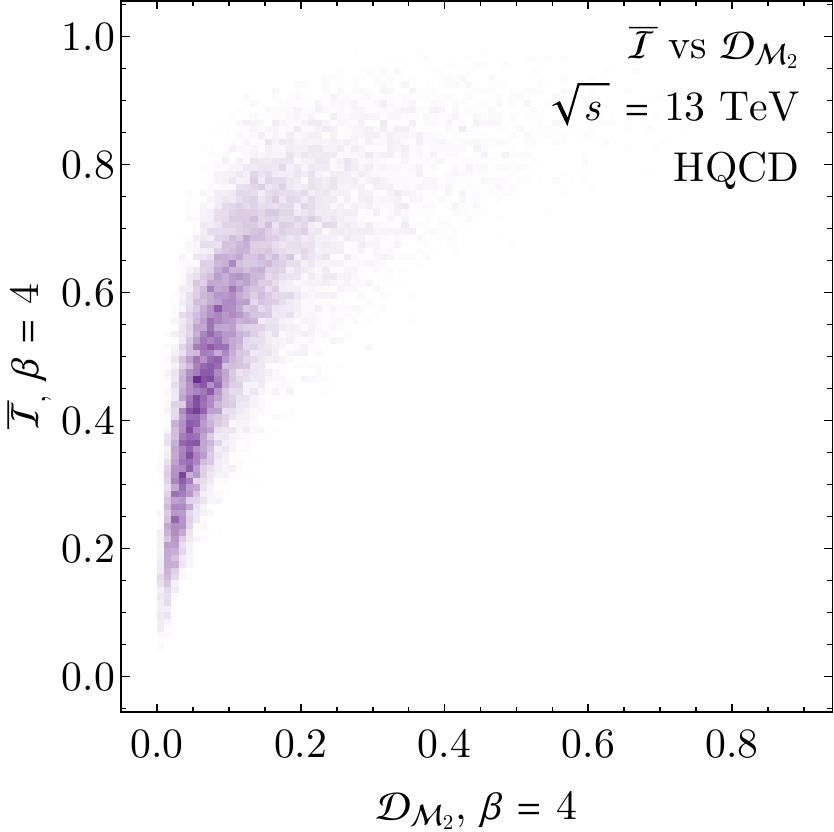}} \quad
\subfigure[]{\includegraphics[width=0.22\textwidth]
{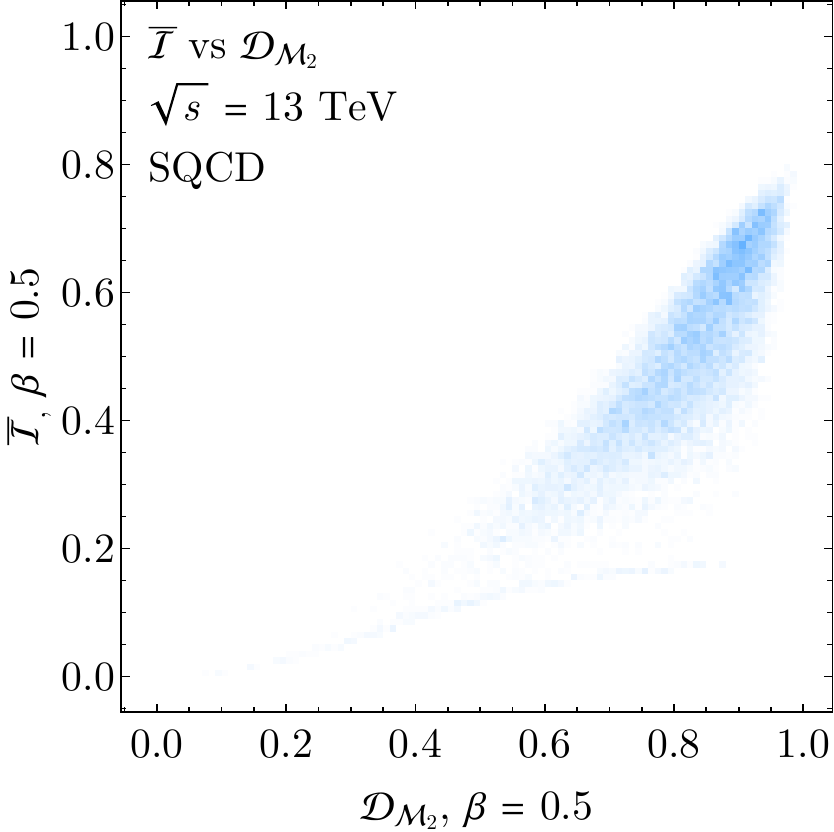}} \quad
\subfigure[]{\includegraphics[width=0.22\textwidth]
{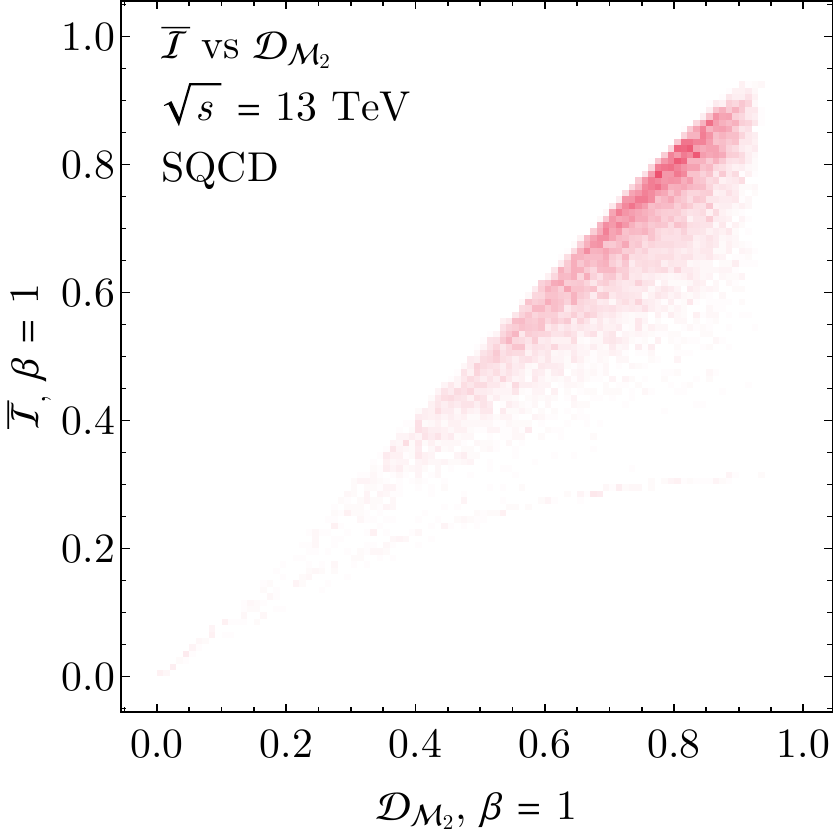}} \quad
\subfigure[]{\includegraphics[width=0.22\textwidth]
{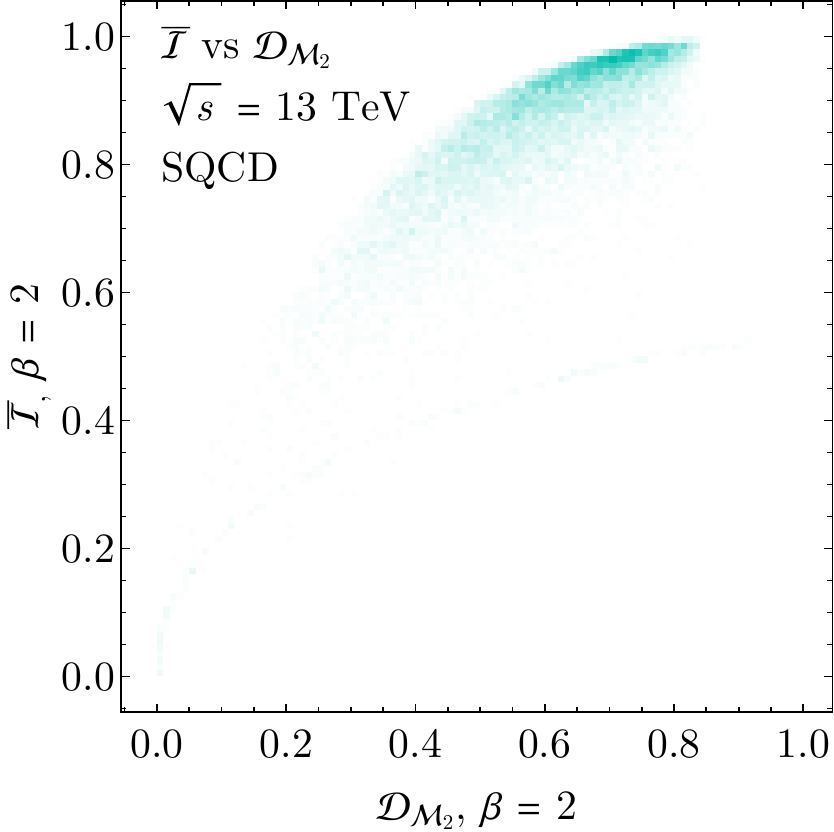}} \quad
\subfigure[]{\includegraphics[width=0.22\textwidth]
{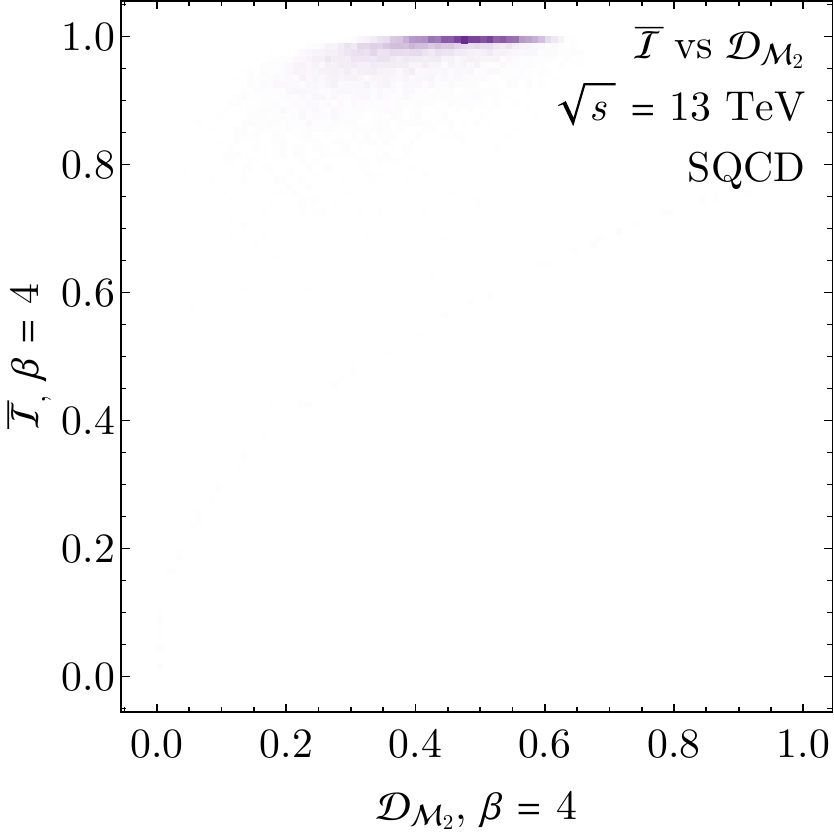}}
\caption{All correlations between \evIso~and \evCol~for HQCD (top) and SCQD (bottom). 
All distributions show some width, mostly for the extremal values of $\beta$, indicating that these observables probe distinct features of the events.}
\label{fig:obsCorr}
\end{figure}

Now let's fix the value of $\beta$ and see how correlated event isotropy and dipole distance are for each physics sample.
The results are shown in Fig.~\ref{fig:obsCorr}. 
We see that for all $\beta$s and QCD samples, the observables are never exactly correlated. 
Some amount of correlation is expected as both observables indicate the underlying event shape, but the value of one observable is not predictive of the other. 

The spread of the 2d correlations in the observables suggests that discrimination of datasets using the 2d plane could be more effective than just the 1d distribution.
The specific choice of $\beta$ that best differentiates the sample is subject to the underlying physics of interest and should be explored for the particular analysis.

\subsection{Particle and Jet Multiplicity}
\begin{figure}[htbp]
\centering
\subfigure[]{\includegraphics[width=0.36\textwidth]
{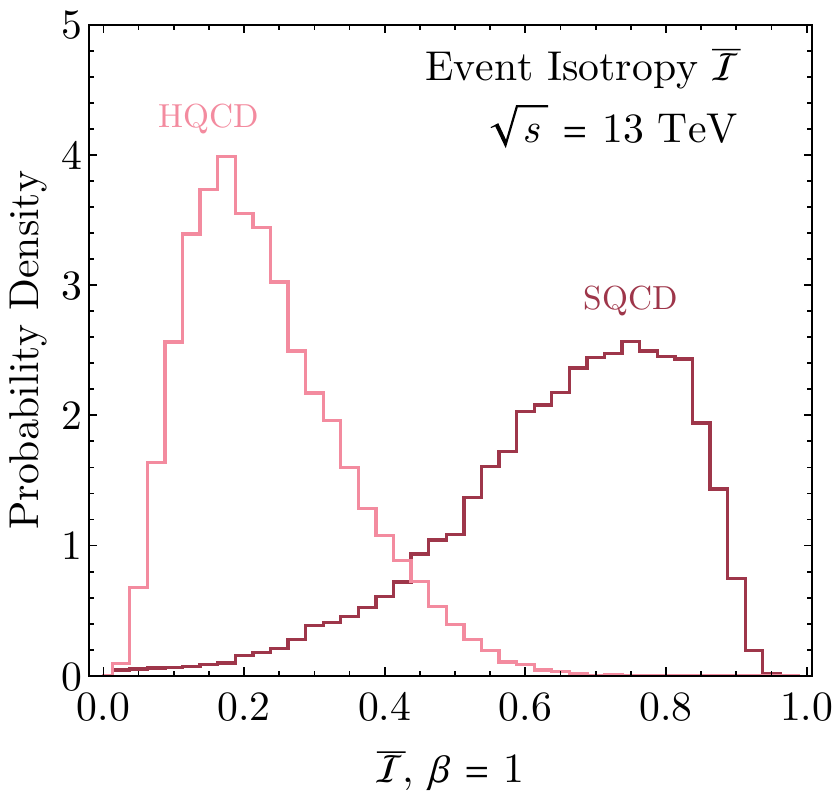}} \qquad
\subfigure[]{\includegraphics[width=0.395\textwidth]
{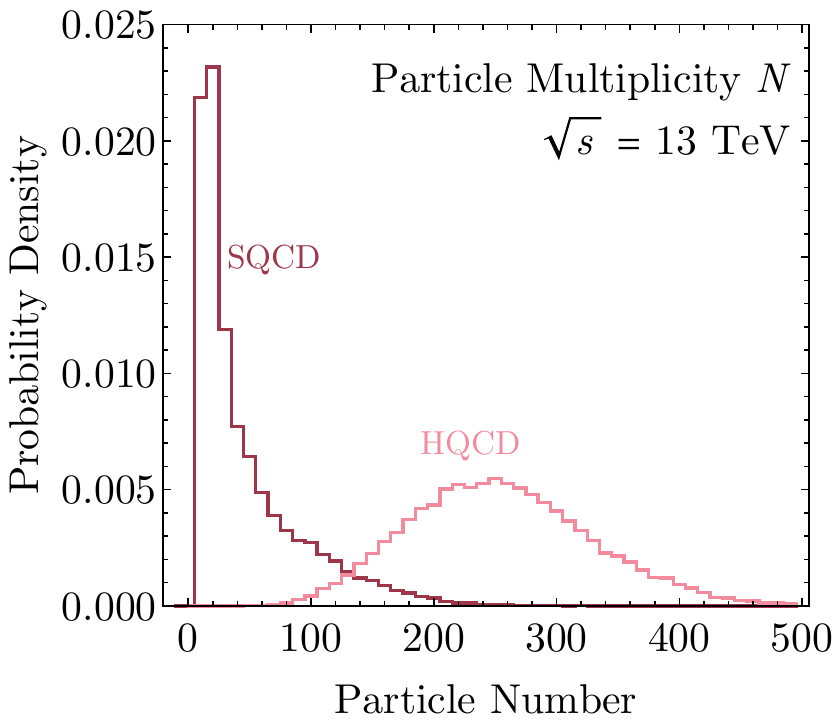}} \quad
\subfigure[]{\includegraphics[width=0.5\textwidth]{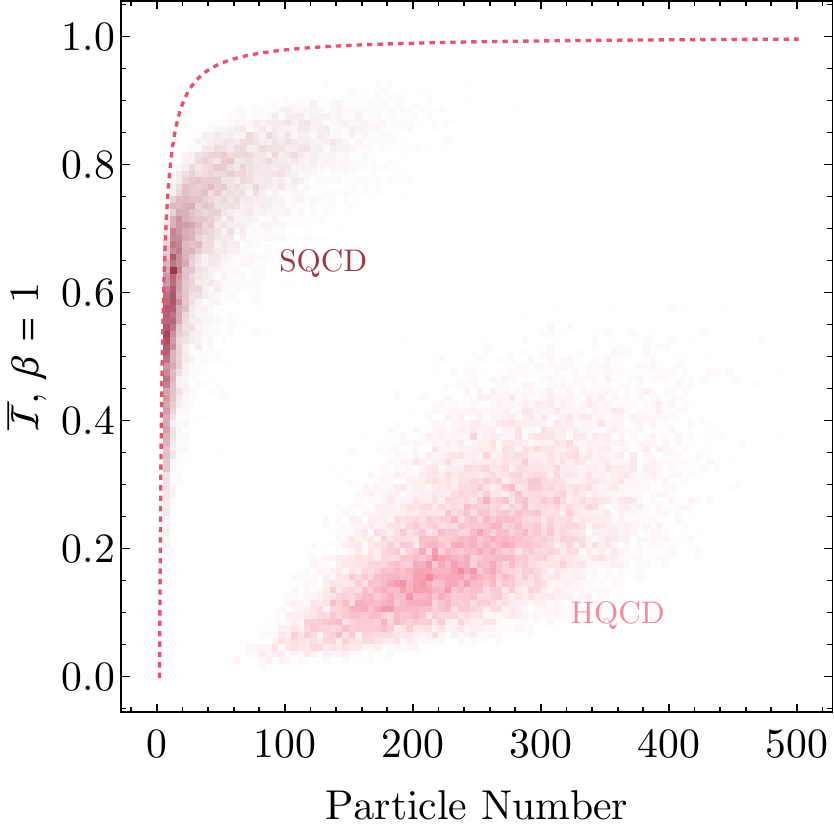}}
\caption{We plot again the distributions in \evIso~from Fig.~\ref{fig:1dbeta1} (a) and in particle number (b), then the combined distribution in (c). 
We see that plotting these observable together completely separates the two samples.}
\label{fig:mult2d}
\end{figure}

A simple observable that is often used to quantify event shapes is particle or jet multiplicity. 
What we will show in this section how these multiplicity measures correlate with manifold distances. 
As stated in Section~\ref{sec:selection}, all particles except neutrinos with $\pt{}>500$ MeV that are within $|y|<4.9$ are counted as part of the total particle multiplicity.
Jets must have $\pt{}>60$ GeV and be within $|y|<4.5$ to be counted in the jet multiplicity.
Often it is naively assumed that small particle number should correlate with collimated underlying physics, and that high multiplicity is needed for an event to look quasi-isotropic.
We see how this is not strictly true in Fig.~\ref{fig:mult2d}. 

When plotting the event isotropy and particle multiplicity, we find that the SQCD and HQCD samples are entirely separated. 
Due to the $p_T$ cut on the particles, the SQCD sample tends to have fewer particles in the final state, but still nearly saturates how isotropic the event could be as measured by event isotropy. 
Meanwhile, the HQCD sample is nowhere near the theoretical limit as low multiplicity events look very non-isotropic as predicted, and only some high-multiplicity events in the MC sample begin to reach beyond the dijet regime. 

\begin{figure}[htbp]
\centering
\includegraphics[width=0.5\textwidth]{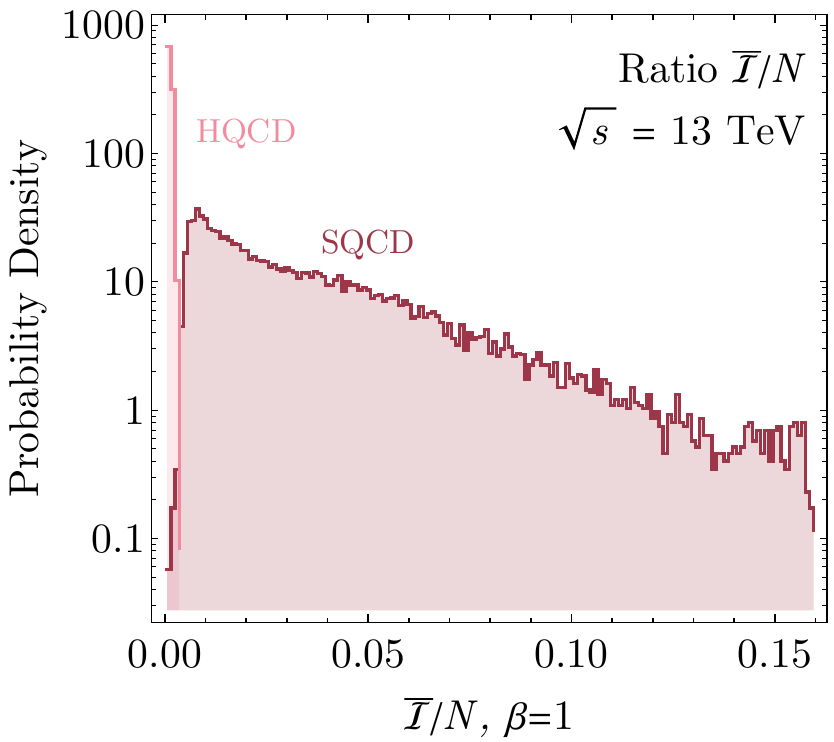}
\caption{The HQCD and SQCD spectra of the ratio between event isotropy \evIso~and particle number $N$.}
\label{fig:ratio2}
\end{figure}

The degree to which these samples can be separated in 2d parameter space is a strong argument for increased use of multi-differential analysis, especially for exploratory studies where the QCD background can be readily identified and removed.
We can even project this 2d distribution back into 1d space by defining a function of the observables $f(\bar{\mathcal{I}}, N)$ that acts as a more powerful discriminant than either variable separately.
For these samples, a very simple function 
\begin{equation}
f(\bar{\mathcal{I}}, N) = \frac{\bar{\mathcal{I}}}{N}
\end{equation}
that is just the ratio of the event isotropy \evIso~to the number of particles $N$ is a highly efficient measure.
As shown in Fig.~\ref{fig:ratio2}, we see that the spectra of the HQCD and SQCD sample are nearly completely separated, with an AUC value $>$ 0.99.
We tested similar measures for the ratio and product of \evIso~and~\evCol, but improvements to the AUC were negligible. 
A more complicated observable (\emph{e.g.} one constructed with machine learning techniques~\cite{hepmllivingreview}) may still exist.
Generally, when observables can probe different features of events, a new function with the observables as inputs of can serve as a highly efficient characterization measure~\cite{Thaler:2010tr,Moult:2016cvt,Komiske:2017aww}.

\begin{figure}[htbp]
\centering
\subfigure[]{\includegraphics[width=0.45\textwidth]
{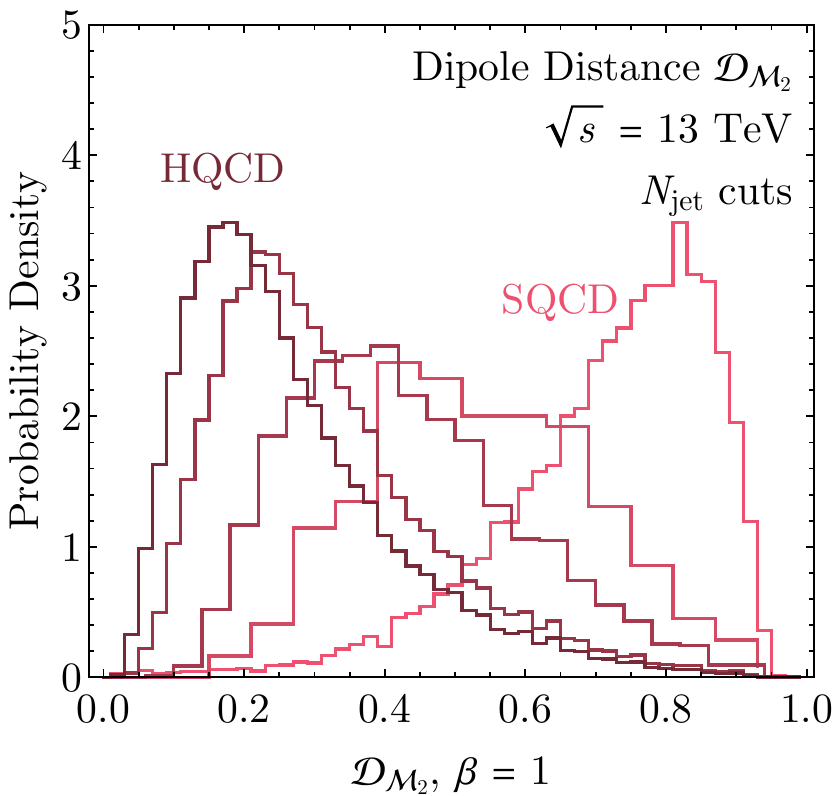}} \qquad
\subfigure[]{\includegraphics[width=0.45\textwidth]
{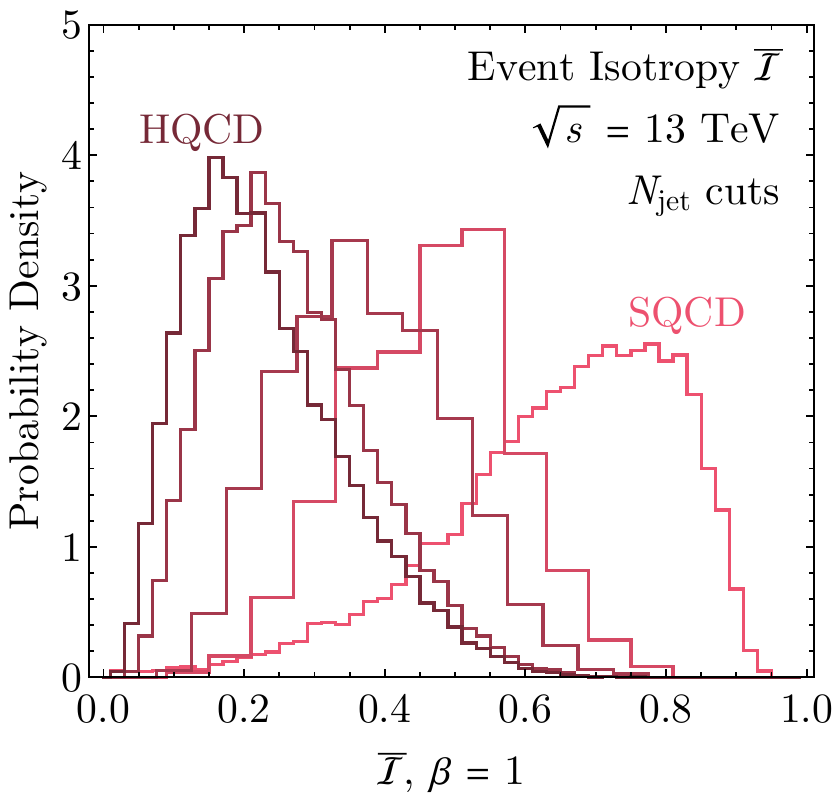}} \quad
\caption{Distributions of the manifold distances for both HQCD and SCQD. The jet cuts are $N_\text{jet} \geq$ 3, 5, 7, and then no jet clustering is applied to SQCD. 
We see the distributions of HQCD shifts towards the SQCD peak as higher jet number cuts are applied.}
\label{fig:jetMult}
\end{figure}

Alternatively, jet multiplicity reflects a more global picture of the radiation pattern as jets have a finite size radius. 
Thus an event with large jet multiplicity suggests that radiation has gone in several distinct directions, whereas a high-multiplicity event could indicate only that there are many collimated particles in a single jet.
To study this correlation, we plot the distribution of the HQCD sample with $N_j \geq$ 2, 3, 5, and 7 in Fig.~\ref{fig:jetMult} along with the SQCD sample for reference.
We see that jet multiplicity tracks the global features of the event more reliably, as the distribution in both event-shape observables shifts towards the SQCD sample, reinforcing one's intuition that higher jet-multiplicity indicates a more isotropic event.

Ultimately, we find that jet multiplicity can be a good indicator for quasi-isotropic radiation patterns, but particle multiplicity is highly decorrelated and can instead be thought of as a partially-correlated, but distinct, handle on the radiation pattern.

\subsection{Hadronization Effects}
Thus far in our analysis, we have focused exclusively on hadronic information. 
In this section we explore how the manifold distances are sensitive to hadronization effects. 
We only consider the HQCD sample as this study is particularly relevant in the context of non-perturbative corrections for jet and event-shape cross-section predictions.

To compare how different the distribution of observables is at hadron- or parton-level, we plot the ratio of the manifold distance computed with hadrons to that of partons. 
If the distribution of the ratio is narrowly peaked at 1, then the partonic vs. hadronic distributions are similar and hadronization does not have a large effect on on the observable.
However, if the distribution is wide, then the observable is significantly different before and after hadronization.

\begin{figure}[htbp]
\centering
{\includegraphics[width=0.45\textwidth]{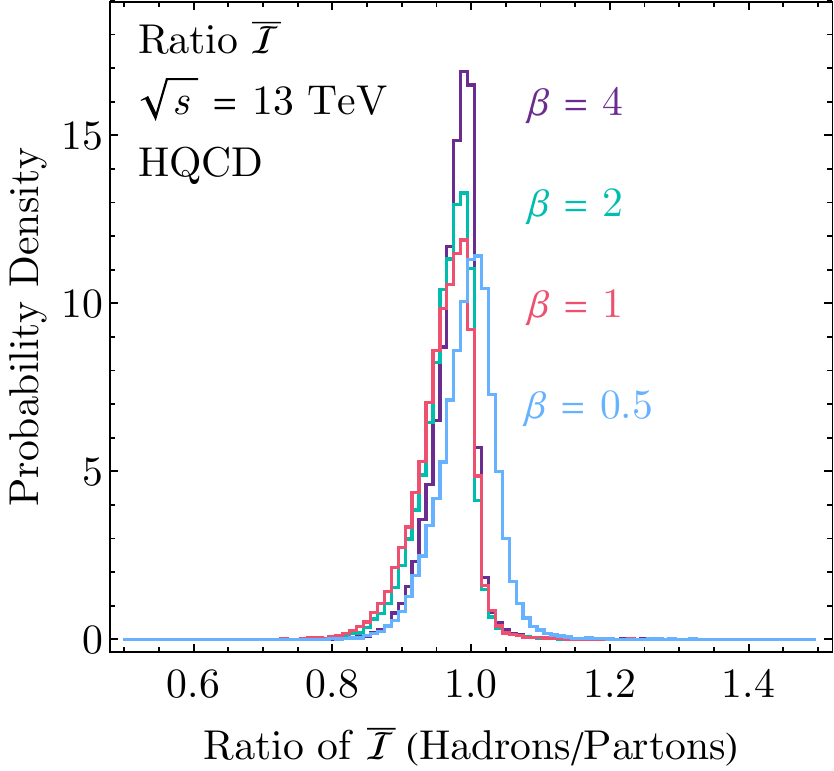}} \quad
{\includegraphics[width=0.45\textwidth]{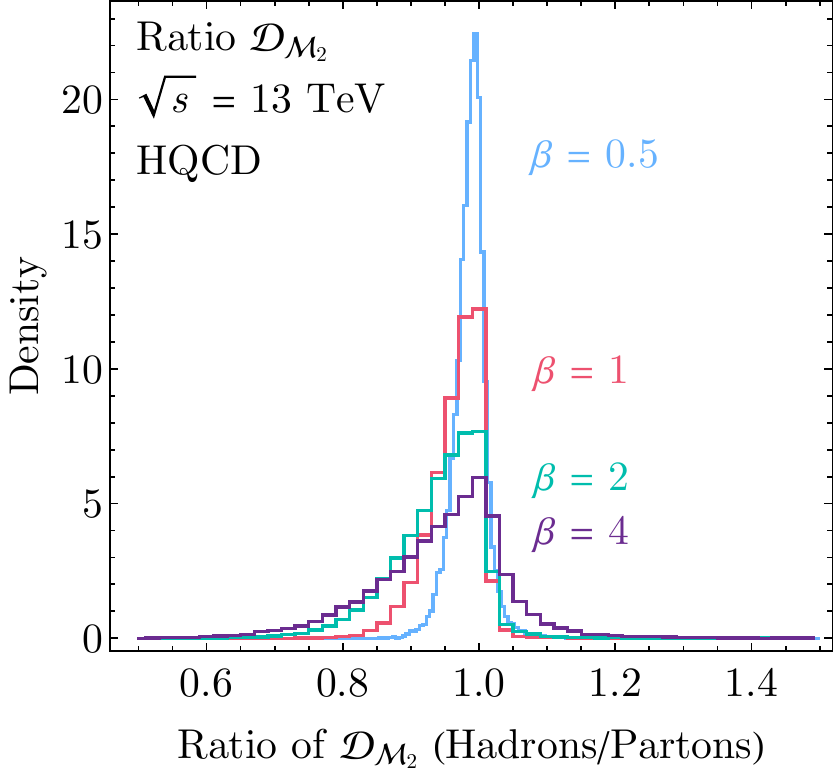}}
\caption{Ratio of manifold distances calculated with hadrons to that calculated with partons. 
Note that ordering of the widths is inverted for \evIso~and \evCol, but generally is less strongly peaked for \evCol, except in the case of high $\beta$.
We anticipate this behavior, as the underlying geometry is more closely related to the dijet reference geometry.}
\label{fig:ratio}
\end{figure}

The results for both event isotropy and dipole distance are shown in Fig.~\ref{fig:ratio}.
The effect of the choice of $\beta$ for event isotropy behaves inversely to that for dipole distance. 
Generally; however, the effect of changing $\beta$ is less significant for event isotropy. 
As $\beta$ increases, the weight assigned to small inter-particle distances decreases, and the more similar the hadron- and parton-level pictures appear. 
In the opposite regime, using dipole distance, we see that larger $\beta$ values correspond to larger differences between the hadron- and parton-level distributions. 
Again, as more particles will have to move a large distance to map on to the reference geometry of the dipole, we expect the distance measure that weights large distance more heavily (large $\beta$) to have a more significant effect on the ratio. 

Sensitivity to hadronization could be a feature or a detriment to the analysis, depending on the question at hand: MC studies may want sensitivity to some aspects of hadronization, while generic new physics searches may not~\cite{Cohen:2023mya}.
We therefore recommend that for hadronization \textit{sensitive} analyses, one should use large $\beta$ for dipole distance and small $\beta$ for event isotropy, whereas the opposite prescription should be applied for hadronization \textit{insensitive} analyses.

\FloatBarrier
\section{Concluding remarks}\label{sec:conclusions}
In this paper we have explored the phenomenology and efficacy of \emph{manifold distances} for several physics scenarios. 
Here we summarize the general trends and behaviors of these observables in order for the reader to choose what observable works best for the analysis at hand.
\begin{itemize}
\item \textit{In general, the manifold distances have the most dynamic range when the underlying physics process is well-approximated by the reference geometry.}
The choice of $\beta$ that exhibits the most range is also dependent on the geometry: small $\beta$ is sensitive to short distances like for a uniform sample, whereas large $\beta$ is sensitive to larger movements, like those often encountered in the dijet sample.
This is not a strict rule, but a good guideline as place to start when developing a new analysis. 

\item \textit{Multi-differential distributions in event-shape observables can be powerful discriminators}. While observables tend to exhibit some degree of correlation, using two at once can provide greater separation between samples than just one.
Using multi-differential analysis is an (underutilized) handle for understanding the flow of hadronic energy, and therefore the underlying physical process---more complementary probes can paint a more complete picture. 
We encourage the community to consider 2d correlations and construct functions of two or more observables for more powerful discrimination prospects.
\item \textit{The degree to which hadronization effects influence the analysis can be tuned using manifold distances.} We have shown ways in which the distributions can be mostly insensitive, and how we can amplify differences due to non-perturbative contributions. As one could envision studies conducted for both scenarios, it is useful to consider how this effect could be included or safely neglected.
\end{itemize}

\noindent
Ultimately, we want to emphasize that we can learn from the success of thrust and generalize the methodology to different processes beyond the soft and collinear features of QCD.
Since the computation of manifold distances is customizable, one has much greater freedom to explore new geometries and applications using a self-consistent language. 
As we enter the era of increased data collection rates at the high-luminosity LHC, and continue in an era of uncertainty as to where new dynamics may be found, we need as many well-understood tools as possible to search robustly for hints of BSM phenomena. 
While we have studied only two physical processes here, there is ample opportunity to use these techniques in the context of searches for new physics. \\

\noindent We encourage the community to take advantage of the flexibility of these observables, and to consider additional reference geometries or distance metrics with the principles described in this work to extend our overall toolkit for analysing collider data.
The code used to compute these observables is publicly available at Ref.~\cite{eventiso}.

\section*{Acknowledgments}
We would like to thank R. Gambhir, G. Kasieczka, K. W. Ho, S. Rappoccio, J. Roloff and J. Thaler for useful discussions.
In particular, we thank M. Feickert for both insight and technical contributions during an early phase of this project.
C.~Cesarotti is supported by the U.S. Department of Energy (DOE) Office of High Energy Physics under Grant Contract No. DESC0012567.
M.~LeBlanc gratefully acknowledges support from the Brown University Department of Physics, and the hospitality of the MIT Center for Theoretical Physics and IAIFI during the final stages of this project.

\appendix
\section{Comparison of Thrust with Dipole Distance}
\label{app:thrust}
The ability to calculate thrust with the EMD was first considered in Ref.~\cite{Komiske:2020qhg}. 
Thrust, which is usually defined as 
\begin{equation}
T \equiv \max_{\hat{n}} \frac{\sum_i |\vec{p}_i\cdot\hat{n}|}{\sum_i |\vec{p}_i|}
\end{equation}
can be rewritten in an equivalent form (for massless particles) as 
\begin{equation}
T \sim \min_{\hat{n}} \frac{\sum_i E_i(1-|\cos\theta_i|)}{E_\text{tot}}.
\end{equation}
From Eqs.~\ref{eq:engNorm} and~\ref{eq:distDef}, we see that this is nearly the definition of the EMD to the dipole event. 
Indeed, the authors of Ref.~\cite{Komiske:2020qhg} note that important distinction is that while thrust is an EMD computation to a back-to-back two particle manifold, the reference geometry is not necessarily balanced, i.e. the two particles could have different energies. 

\begin{figure}
\begin{center}
\includegraphics[width=0.3\textwidth]{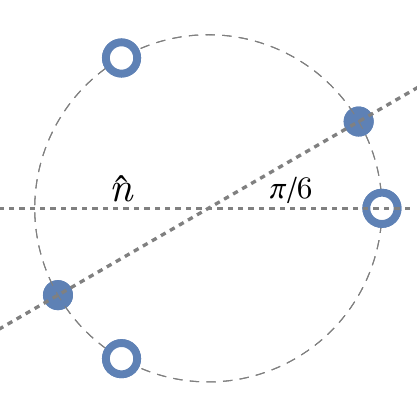}
\caption{An illustration of the difference in computing \evCol~to the two-particle configuration and thrust. 
The thrust axis is as marked, and the two-particle configuration that minimizes the distance is shown on the plot at an angle of $\pi/6$.}
\label{fig:thrustVemd}
\end{center}
\end{figure}

To illustrate how these quantities are different, let's consider a simple event: the three-prong configuration. 
When computing thrust, the thrust axis is aligned with any of the three prongs.
However, the dipole distance \evCol~is minimized for a two-particle event at an angle $\pi/6$ displaced from any of the three particles (see Fig.~\ref{fig:thrustVemd}).

We can understand the difference with geometry. 
While thrust is computing the minimum distance to an \textit{axis}, the EMD is computing the distance between individual \textit{points} in sets. 
Because of this, thrust is not the manifold distance to back-to-back particle events with equal energy, but rather with potentially unbalanced energy, and is therefore in a different class of observables than manifold distances. 

\clearpage
\bibliographystyle{utphys}
\bibliography{ref,ATLAS,CMS,PubNotes,ConfNotes,ATLAS-useful}

\end{document}